\documentclass[aps,
	twocolumn,
	altaffilletter,
	nolongbibliography,
	numerical,
	flushbottom,
	secnumarabic,
	pra,
	superscriptaddress,
	floatfix,
	10pt]{revtex4-2}

\usepackage[T1]{fontenc}

\usepackage{booktabs}
\usepackage{newtxtext,newtxmath}
\usepackage{graphicx}
\usepackage{braket}
\usepackage{hyperref}
\hypersetup{%
	colorlinks=true, 
	breaklinks=true,
	urlcolor=blue, 
	linkcolor=blue, 
	citecolor=blue,
}

\begin{document}

\title{Convergence of Digitized-Counterdiabatic QAOA: circuit depth versus free parameters}%

\author{Mara Vizzuso}
\affiliation{Dipartimento di Fisica ``E. Pancini'', Universit\`a degli Studi di Napoli ``Federico II'', Complesso Universitario M. S. Angelo, via Cintia 21, 80126, Napoli, Italy}

\author{Gianluca Passarelli}
\altaffiliation[Present affiliation: ]{Dipartimento di Fisica ``E. Pancini'', Universit\`a degli Studi di Napoli ``Federico II'', Complesso Universitario M. S. Angelo, via Cintia 21, 80126, Napoli, Italy}
\affiliation{CNR-SPIN, c/o Complesso Universitario M. S. Angelo, via Cintia 21, 80126, Napoli, Italy}

\author{Giovanni Cantele}
\affiliation{CNR-SPIN, c/o Complesso Universitario M. S. Angelo, via Cintia 21, 80126, Napoli, Italy}

\author{Procolo Lucignano}
\email{procolo.lucignano@unina.it}
\affiliation{Dipartimento di Fisica ``E. Pancini'', Universit\`a degli Studi di Napoli ``Federico II'', Complesso Universitario M. S. Angelo, via Cintia 21, 80126, Napoli, Italy}

\date{\today}

\begin{abstract}
Recently, Digitized-Counterdiabatic (CD) Quantum Approximate Optimization Algorithm (QAOA) has been proposed to make QAOA converge to the solution of an optimization problem in fewer steps, inspired by Trotterized counterdiabatic driving in continuous-time quantum annealing. In this paper, we critically revisit this approach by focusing on the paradigmatic weighted and unweighted 
one-dimensional MaxCut problem. We study two variants of QAOA with first and second-order CD corrections. Our results show that, indeed, higher order CD corrections allow for a quicker convergence to the exact solution of the problem at hand by increasing the complexity of the variational cost function. Remarkably, however, the total number of free parameters needed to achieve this result is independent of the particular QAOA variant analyzed {for the problems considered}.
\end{abstract}

\maketitle

\section{\label{sec:Intro} Introduction}

In the last decade, variational quantum eigensolvers such as the quantum approximate optimization algorithm (QAOA)~\cite{farhi2014quantum} have sparked the interest of the scientific community~\cite{Arufe2023, sureshbabu2023parameter, stechly2023connecting, bornens2023variational, wan2023hybrid, vikstal2023quantum,Zawalska2023,Torta2023, wauters2020polynomial, Xue_2021, Streif_2020, zhu2022, pan2022, Dupont2022,10.1088/1367-2630/ace547}, proving to be valuable tools in tackling the Hamiltonian-to-ground-state problem~\cite{farhi2014quantum,cerezo:2021,mbeng2019quantum,Dlaska2022}. 

Given a Hamiltonian $H_T$, which encodes the solution of a hard binary optimization task~\cite{lucas:2014}, the main idea is to start from a trivial $N$-qubit wavefunction and let it evolve by repeatedly applying two kinds of unitaries: a simple one, generated by a transverse-field mixer Hamiltonian $H_X$, providing quantum fluctuations independently to each particle, and a hard one, generated by the target many-body interacting Hamiltonian $H_T$. 

To start, a quantum processor is employed to natively embed a parametric quantum circuit where each unitary acts for a certain time, to be optimized minimizing a cost function. The “time” parameters are known as angles in the QAOA paradigm, and are optimized using traditional computational routines on a classical computer. The cost function is the expectation of $H_T$ over the circuit state. In QAOA, the quantum processing unit is used as a hardware accelerator to speed-up the computation of the cost function. 
Therefore, the QAOA is a hybrid classical-quantum algorithm, and it has been recently implemented on different quantum platforms \cite{Proietti_2022, shaydulin2023, Weidenfeller2022scalingofquantum, Kim2023}. 

Given enough expressiveness, the QAOA circuit represents an approximation of the target ground state that becomes better and better the more parameters are employed. It has been proven that QAOA is computationally universal~\cite{lloyd:2018,morales:2020} and equivalent, in the infinite-depth limit $p\to\infty$ (with $p$ the number of QAOA steps), to adiabatic quantum computing (AQC)~\cite{kadowaki:1998,santoro-martonak,farhi:2000,albash:2018,mbeng2019optimal}. In fact, QAOA was originally proposed as a Trotterized, digital version of AQC where the Trotter error arising from the noncommuting operators $H_T$ and $H_X$ is zero when the integration time step is infinitesimal. Thus, the adiabatic theorem sets bounds on the performance of QAOA the large $p$~\cite{Born1928} regime. 
However, when the number of parameters grows, classical optimization routines often struggle to find good minima due to the appearance of exponentially vanishing gradients of the cost function, a phenomenon known as barren plateaus~\cite{biamonte:reachability-qaoa}. Even though avoiding these plateaus might prove to be impossible in the case of spin glass Hamiltonians~\cite{Larocca2022diagnosingbarren, Hashim2022, Basso2022, Streif_2020}, smart choices of the initial condition of the minimization process~\cite{LeoZhou2020, mbeng:avoiding-barren-plateaus, cain2022qaoa, tate2022warmstarted, Bartschi2020} or artificial intelligence~\cite{stein2023evidence, Ha2023, Choi2020, Shaydulin:2020jdg} can help circumvent this issue.

Moreover, in the era governed of noisy intermediate-scale quantum devices~\cite{Bharti2022}, low-depth QAOA circuits are much more practical since they can be physically implemented in real devices and are more robust against noise~\cite{weidinger2023error,Wauters2020}. For small values of $p$, the analogy between QAOA and digitized AQC breaks down and QAOA has to be considered as a completely separate heuristic optimization method. 

{Recently, many QAOA variants have emerged to enhance its performance~\cite{Hadfield2019, Bartschi2020,Villalba-Diez2022,Golden2021,Fuchs2022,egger2021,Yoshioka2023,Wurtz2021,cadavid2023}. A non-exhaustive list of these variants includes: QAOA$+$~\cite{Chalupnik2022}, which enhances the conventional QAOA by incorporating an extra problem-independent layer with multiple parameters; adaptive-bias QAOA~\cite{Yu2022}, which adds local fields to the QAOA operators to decrease computation time; adaptive QAOA~\cite{zhu2022}, a version of QAOA that iteratively selects mixers based on a systematic gradient criterion; recursive-QAOA~\cite{Bravyi2020}, which aims to reduce the problem size by eliminating unnecessary qubits following a non-local scheme. This proliferation of variants highlights the ongoing effort to refine and extend QAOA for improved adaptability and efficiency in diverse quantum computing applications.}  

In addition, several strategies partly inspired by diabatic quantum computation have been ported to the QAOA language to further enhance its performance~\cite{An2022,farhi:quantum-supremacy-qaoa}. Successful results have been achieved, for instance, by adding unitaries generated by a pool of local operators~\cite{Chandarana:2022}, adding $XX$-$YY$ interactions~\cite{Chai:2022}, using reinforcement learning~\cite{wauters:prr2020,bukov:2021}, qubit-dependent angles like in multi-angle QAOA~\cite{herrman:multi-angle-qaoa}, or, ultimately, taking into account Trotterization errors via next-order Baker-Hausdorff-Campbell (BHC) expansion~\cite{Wurtz2022counterdiabaticity, chandarana2022digitizedcounterdiabatic}. The latter scheme has been named digitized-counterdiabatic QAOA since it exploits operators arising from the Lie algebra of $H_X$ and $H_T$ to improve the algorithm, similarly to the nested commutator expansion of the adiabatic gauge potential used for shortcuts to adiabaticity in the adiabatic computing paradigm~\cite{Claeeys:2019}. This approach resulted in a recent experimental demonstration of quantum integer factorization using a trapped-ion quantum processor~\cite{hegade:qaoa-cd-factorization}, suggesting that the key to unlock quantum advantage may rely on these hybrid schemes including corrections beyond first-order. Yet, there is still no clear understanding of these counterdiabatic corrections in QAOA and many open questions remain. The existing literature to this date mostly focused on first-order counterdiabatic corrections (referred to as QAOA-CD in the following) \cite{Hegade2022}. Here, we will make a step forward and also discuss the second-order correction to the BHC formula (QAOA-2CD). We will give more details in the following sections, where we numerically address some of the open points of counterdiabatic QAOA, focusing our attention on the paradigmatic MaxCut problem for which QAOA had been originally proposed~\cite{farhi2014quantum}. 

We consider  one-dimensional (1D) chains of $N$ spins with uniform or random interaction strengths, which can be easily mapped onto the 1D Ising model. Thanks to the Jordan-Wigner transformations, we can switch to a fermionic representation of the system Hamiltonian and run simulations with a reasonable computational effort [$O(N)$] even in the presence of disorder~\cite{mbeng2020quantum}.

The rest of this paper is organized as follows.  We discuss the properties of the MaxCut Hamiltonian in Sec.~\ref{sec:model} and
review QAOA and its counterdiabatic corrections in Sec.~\ref{sec:qaoa}. In Sec.~\ref{sec:results}, we show the results of our numerical analysis for selected instances. The main finding here is that, even though counterdiabatic corrections allow to decrease the circuit depth $p$ if compared to standard QAOA, the number of angles necessary for convergence to a given tolerance remains the same. We finally draw our conclusions in Sec.~\ref{sec:conclusions}.

\section{MaxCut Model} \label{sec:model}

In graph theory, a graph is a set of vertices connected by edges. Given a graph, the maximum cut is a line that cuts the largest number of edges between two disjoint sets of vertices, such that the cut size (defined as the number of broken edges) is at least the size of any other cut. Finding this maximum cut is the NP-hard problem known as MaxCut.
This problem can be expressed in a quadratic unconstrained binary optimization (QUBO) form and mapped to an Ising Hamiltonian~\cite{lucas:2014}. Thus, finding the maximum cut is equivalent to finding the ground state of the corresponding Ising Hamiltonian. This problem can be tackled using techniques from statistical and quantum physics, such as quantum and simulated annealing~\cite{kadowaki:1998,albash:2018,simulated-annealing-book,phegde:ga,gpassarelli:qa1,gpassarelli:qa2,gpassarelli:qa3,gpassarelli:qa4,gpassarelli:qa5,gpassarelli:qa6,gpassarelli:qa7,10.1088/1367-2630/ace547}, or QAOA. In what follows, we will focus on the latter.

We start defining the antiferromagnetic Ising model Hamiltonian,
\begin{equation}
    H_T = J \sum_{ \langle i,j\rangle} ^N \sigma_i^Z \sigma_j^Z,
    \label{htarget}
\end{equation}
where angular brackets denote sites connected by a graph edge, $\sigma^Z_j$ are Pauli matrices, and $N$ is the number of vertices. We will express energies in units of the  (positive) exchange coupling $J$ and omit the latter, when unnecessary, for ease of notation. The minimum energy configuration of Eq.~\eqref{htarget} corresponds to the maximum cut of the assigned graph, with the 
cutting line separating positive from negative spins. 
This model can be generalized assigning different weights to the various edges, which is usually named weighted MaxCut problem, described by the Hamiltonian
\begin{equation}
    H_T = \sum_{ \langle i,j\rangle} ^N J_{i,j}  \sigma_i^Z \sigma_j^Z.
    \label{htarget-weighted}
\end{equation} 

Depending on the specific choice of the couplings $J_{i,j}$, and on the graph structure, we can address several models. In this manuscript, we mostly focus on three specific cases: (i) the uniform antiferromagnetic Ising chain with periodic boundary conditions (PBC), known as ring of disagrees since spins are anti-aligned in the ground state; (ii) the uniform antiferromagnetic Ising chain with open boundary conditions (OBC); (iii) the Ising chain with OBC and random couplings. For all these models we analyze chains with $N=10,16,20$ spins.

\subsection{Ring of disagrees}

As mentioned, the ring of disagrees is a one-dimensional chain of 
spin-$1/2$ particles with nearest-neighbor antiferromagnetic couplings. For this type of system, Eq.~\eqref{htarget} becomes
\begin{equation}
    H_T =  \sum_{i=1}^N \sigma_i^Z \sigma_{i+1}^Z,
    \label{HT-ring}
\end{equation}
where the couplings between nearest neighbours are  set to $J_{i,i+1}=1 \; \forall i$. Periodic boundary conditions imply that  $\sigma_{i+N}^Z=\sigma_i^Z\; \forall i$. For this specific model, analytical results regarding standard QAOA are available~\cite{farhi2014quantum,rieffel:2018}. Thus, we will use this model as a test bed for our analysis.

\subsection{Open chain}\label{open chain}

For the open chain with uniform couplings $J_i = J = 1\;\forall i$, Eq.~\eqref{htarget} becomes
\begin{equation}
    H_T=\sum_{i=1}^{N-1}\sigma_i^Z\sigma_{i+1}^Z.
    \label{h-target-open}
\end{equation}
In this case, there are no analytical  predictions as opposed to the PBCs case, and some of the considerations that are discussed in the literature regarding the ring of disagrees of  Eq.~\eqref{HT-ring} are not applicable to the open chain.

\subsection{Random couplings}

Another model studied in this paper is the open chain with random coupling energies, that we choose to be uniformly distributed [$J_i = \mathcal{U}([-1, 1])$, where $\mathcal{U}$ indicates an uniform distribution in the given range]. $H_T$ becomes
\begin{equation}
    H_T=\sum_{i=1}^{N-1}J_i\sigma_i^Z\sigma_{i+1}^Z.
    \label{h-target-random}
\end{equation}
For each chain we generate $M=20$ random instances and repeat our analysis for all of them.  For all these models our goal is to  compare QAOA, QAOA-CD and QAOA-2CD to figure out which algorithm is the most advantageous.

\section{QAOA and its variants}
\label{sec:qaoa}
In this section we briefly outline the main concepts and equations underlying the QAOA algorithm and its variants used throughout the paper.

Conventional  QAOA \cite{farhi2014quantum} consists in $p$ repeated applications of two unitaries defined as $U_H(\lambda) = e^{-i \lambda H}$, where $\lambda$ is a parameter, to be optimized, and $H$ is either the target Hamiltonian $H_T$ or  a the transverse-field Hamiltonian defined as
\begin{equation}
    H_X = \sum_{i=1}^N \sigma_i^X,
    \label{hx}
\end{equation}
where $\sigma^X_i$ is the Pauli matrix acting on the $i^\text{th}$ qubit. 
We choose as starting state the ground state of $H_X$, which, in the computational eigenbasis of the Pauli matrices $\sigma^Z_i$, reads
\begin{equation}\label{start_state}
    \ket{0} = \frac{1}{\sqrt{2^N}} \bigotimes_{i=1}^N \left( \ket{\uparrow}_i - \ket{\downarrow}_i \right),
\end{equation}
and a variational wavefunction

\begin{align}
   &\ket{\psi^{(p)}(\vec \gamma,\vec \beta)}= \prod_{k=1}^{p} U^{(k)}(\gamma_k, \beta_k) \ket{0} \notag \\ &\quad= [U_{H_X}(\beta_p)U_{H_T}(\gamma_p)]\cdots [U_{H_X}(\beta_1) U_{H_T}(\gamma_1)]\ket{0},
   \label{psi-qaoa}
\end{align}
where $U^{(k)}(\gamma,\beta) = U_{H_X}(\beta) U_{H_Z}(\gamma)$ and $p$ labels the number of QAOA steps.
The ground-state wavefunction of $H_T$ is then approximated minimizing the energy function
\begin{equation}
 E_p(\vec \gamma, \vec \beta) =  \braket{\psi^{(p)}(\vec \gamma,\vec \beta) | H_T | \psi^{(p)}(\vec \gamma,\vec \beta)}
 \label{fun-QAOA}
\end{equation}
with respect to the $2p$ variational parameters $\vec \gamma=(\gamma_1,...,\gamma_p)$, $\vec \beta=(\beta_1,...,\beta_p)$:
\begin{equation}
    (\vec \gamma^*, \vec \beta^*) = \arg\min_{\vec \gamma, \vec \beta} E_p(\vec \gamma, \vec \beta).
\end{equation}
In the limit $p\to \infty$,  $E_p(\vec \gamma^*,\vec \beta^*)\to E_\text{GS}$ and $\ket{\psi^{(p)}(\vec \gamma^*,\vec \beta^*)} \to \ket{\psi_\text{GS}}$ where $E_\text{GS}$ and $\ket{\psi_\text{GS}}$ are the true ground state energy and wavefunction of $H_T$, respectively.
In some special cases, it is possible to achieve the exact solution in a finite number of iterations $p$~\cite{farhi2014quantum,rieffel:2018}. Even when this is not the case, often just few iterations allow one to get a good approximate solution, whose accuracy  can be quantified by the residual energy, defined as
\begin{equation}
\varepsilon_\text{res}^p=\frac{\min_{\vec \gamma,\vec \beta} E_p(\vec{\gamma},\vec{\beta})-E_\text{min}}{E_\text{max}-E_\text{min}},
    \label{eres}
\end{equation}
with its straightforward generalizations to QAOA-CD and QAOA-2CD, defined later. Here,  $E_\text{max (min)}$  is the largest (smallest) eigenvalue of $H_T$. As such, $\varepsilon_\text{res}^p$ measures the difference between the minimum energy at the $p$-th step and the
numerically exact ground-state energy, normalized with respect to the full (numerically exact) spectrum width 
$E_\text{max}-E_\text{min}$.

\subsection{QAOA-CD}\label{QAOA-CD}
In this work, the point is modifying the QAOA algorithm to improve its  performances at fixed number of iterations $p$. The main idea is to  include, at each iteration, new unitaries inspired by the missing terms of the BHC expansion. In    particular, the Zassenhaus formula~\cite{casas:zassenhaus} reads
\begin{equation}\label{eq:zassenhaus}
    e^{aX+bY} = e^{aX} e^{bY} \prod_{n=2}^\infty e^{C_n(aX, bY)}
\end{equation}
where $a,b\in\mathbb{R}$ and, for each pair of operators $X$ and $Y$, $C_n(aX, bY)$ are homogeneous Lie polynomials in $X$ and $Y$. For $n=2$ we have
\begin{equation}
    C_2(aX, bY) = \frac{ab}{2}[X, Y]. 
    \label{BCH-2}
\end{equation}
In what we denote as QAOA-CD, we keep the $n = 2$ term in Eq.~\eqref{eq:zassenhaus} and write,
following Eq.~\eqref{BCH-2}, each unitary operator of the algorithm as
\begin{equation}\label{eq:qaoa-cd}
    U^{(k)}(\gamma_k, \beta_k, \alpha_k) = U_{H_X}(\beta_k) U_{H_Z}(\gamma_k) U_\text{CD}(\alpha_k),
\end{equation}
where
\begin{equation}
    U_\text{CD}(\alpha_k) = e^{\alpha_k [H_X,H_T]}.
    \label{UCD}
\end{equation}
The expectation of $H_T$ over the variational wavefunction now depends on $3p$ parameters instead of the $2p$ parameters of regular QAOA. The cost function in this case reads
\begin{equation}\label{eq:cost-QAOA-CD}
    E_p(\vec \gamma, \vec \beta, \vec \alpha) = \braket{\psi^{(p)}(\vec \gamma,\vec \beta,\vec \alpha) | H_T | \psi^{(p)}(\vec \gamma,\vec \beta,\vec \alpha)}.
\end{equation}
At each step $p$, the $3p$ parameters $(\vec{\gamma}^*,\vec{\beta}^*,\vec{\alpha}^*)$ minimize the cost functional of Eq.~\eqref{eq:cost-QAOA-CD} and $\ket{\psi^{(p)}(\vec{\gamma}^*, \vec{\beta}^*,\vec{\alpha}^*)}$ is the approximate ground state.

\subsection{QAOA-2CD}\label{QAOA-2CD}
While the expansion of Eq.~\eqref{eq:zassenhaus}  has recently been  studied~\cite{Chandarana:2022,Wurtz2022counterdiabaticity}, higher order corrections are still unexplored. 
In this paper we want to consider, alongside the operator given by Eq.~\eqref{UCD}, a new operator inspired by the next-order expansion of Eq.~\eqref{eq:zassenhaus}. The Lie polynomial given by the third order of Eq.~\eqref{eq:zassenhaus} is 
\begin{equation}
    C_3(aX, bY) = \frac{a^2b}{6} [X, [ X, Y]] - \frac{ab^2}{3} [Y, [X, Y]].
    \label{secon term}
\end{equation}
We get what we call QAOA-2CD. In particular, we define the unitary operator at $k^\text{th}$ step as
\begin{align}\label{eq:qaoa-2cd}
    &U^{(k)}(\zeta_k, \delta_k, \gamma_k, \beta_k, \alpha_k) \notag \\ &\quad= U_{H_X}(\beta_k) U_{H_Z}(\gamma_k) U_\text{CD}(\alpha_k) U_\text{2CD}(\delta_k, \zeta_k),
\end{align}
where 
\begin{equation}
    U_\text{2CD}(\delta_k, \zeta_k) = e^{i\delta_k [H_X, [H_X, H_T]] - i\zeta_k [H_T, [H_X,H_T]]}.
    \label{eq:u2cd}
\end{equation}
The cost function $E_p(\vec \zeta, \vec \delta, \vec \gamma, \vec \beta, \vec \alpha)$ readily generalizes Eq.~\eqref{eq:cost-QAOA-CD}. Notice that in this case we consider five variational parameters per QAOA-2CD step, hence the total number of free angles in this variant of the algorithm is $5p$.

A question arises on whether the parameters resulting from the $p$-step minimization might be used 
to guess the starting parameters at the the $(p+1)$-th step.
Within all three frameworks, QAOA, QAOA-CD and QAOA-2CD, we adopt the INTERP recipe of 
Ref.~\cite{Zhou2020}. Taking as an example the $\vec\gamma$'s, it can be shown that a good starting
guess at the step $p+1$ is given by the linear interpolation
\begin{equation}
    \left[\vec{\gamma}^\text{guessed}_{(p+1)}\right]_i=\frac{i-1}{p}\left[\vec{\gamma}^*_{(p)}\right]_{i-1}+\frac{p-i+1}{p}\left[\vec{\gamma}^*_{(p)}\right]_i,
    \label{interpolazione}
\end{equation}
where $i$ denotes the $i$-th component of the vector, $i=1,\dots, p+1$, $\vec{\gamma}^*_{p}$ is the $p$-th
component of the $\vec\gamma$ as obtained from the minimization at the $p$-th step. It is assumed that
$\left[\vec{\gamma}^*_{(p)}\right]_0 = \left[\vec{\gamma}^*_{(p)}\right]_{p+1}=0$.
 We will see in the next sections that, while this recipe well performs in most cases,
random chains are more efficiently handled starting from scratch at each
step $p$.

\section{Results} \label{sec:results}
In this section, we critically discuss the main outcomes obtained within the QAOA-CD and QAOA-2CD frameworks.
To analyze the quality of our algorithms, we use as control parameter at the
$p$-th step the residual energy as defined in Eq.~\eqref{eres}.

Before proceeding further, an important remark has to be done as 
far as the counterdiabatic corrections are concerned. The QAOA-CD and
QAOA-2CD cost functions, $E_p(\vec \gamma, \vec \beta, \vec \alpha)$
[Eq.~\eqref{eq:cost-QAOA-CD}] and
$E_p(\vec \zeta, \vec \delta, \vec \gamma, \vec \beta, \vec \alpha)$,
respectively depend on $3p$ and $5p$ parameters. These are
conceived based on the intuition that adding higher order terms in the BHC expansion \cite{math6080135} reduces the Trotter error.
In principle, according to the Zassenhaus formula of Eq.~\eqref{eq:zassenhaus},  $\alpha_k, \delta_k$ and $\zeta_k$ are not free parameters but rather they are related to $\beta_k$  and $\gamma_k$ by the following relations:
\begin{equation}\label{eq:no-free-params}
    \alpha_k = -\beta_k \gamma_k / 2, \quad \delta_k = \beta_k^2 \gamma_k / 6, \quad 
    \zeta_k = \beta_k \gamma_k^2 / 3.
\end{equation}
If these constraints are assumed,
the above-mentioned
cost functions depend, at the $p$-th step, on $2p$ parameters.
The variants of  QAOA-CD and QAOA-2CD, where Eqs.~\eqref{eq:no-free-params}
are taken into account, will be referred to, in the following, as QAOA-CD-$2p$ and QAOA-2CD-$2p$, respectively.

\begin{figure}
    \centering
    \includegraphics[width=0.8\columnwidth]{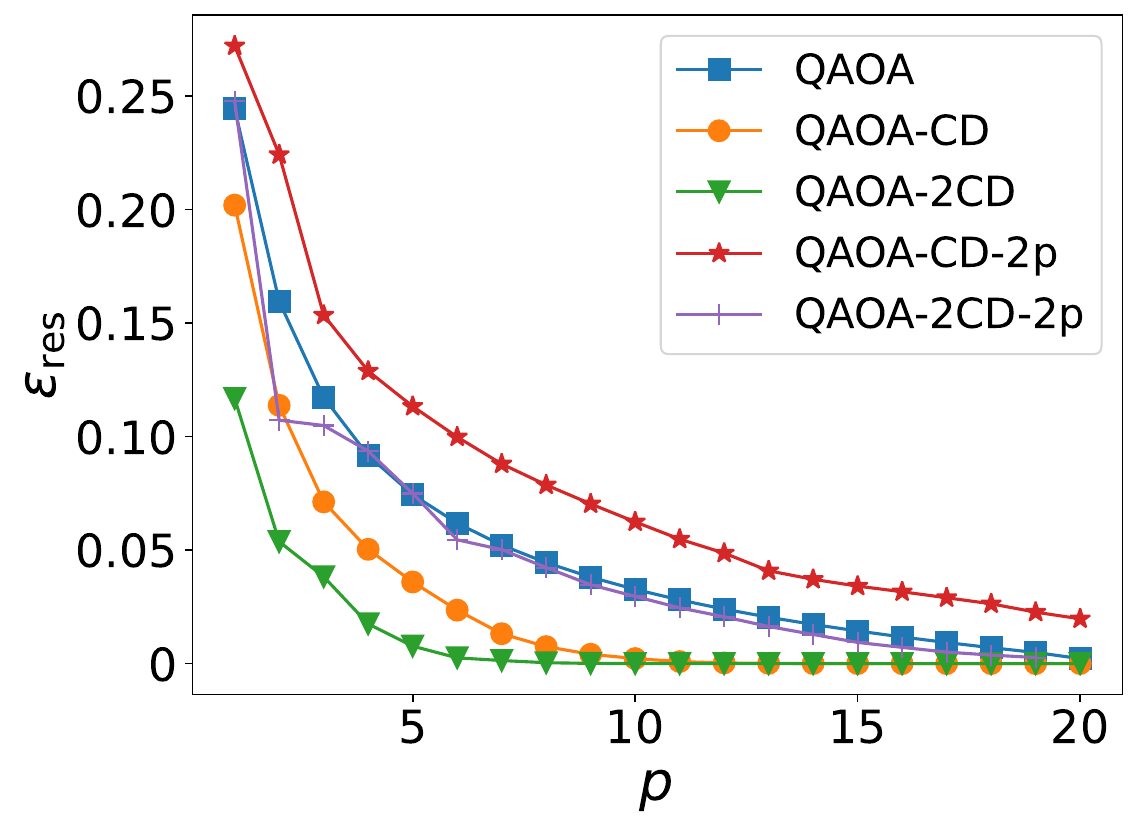}
    \caption{Residual energy for a graph of $N=20$ spins with OBC and uniform couplings. The results of
    the QAOA-CD-$2p$ and QAOA-2CD-$2p$ (depending both on $2p$
    parameters) and of QAOA-CD and QAOA-2CD (depending on $3p$ and $5p$
    parameters, respectively) are shown. The latter two exhibit the best
    performance in the convergence rate towards the ground state and,
    at any $p$, produce lower residual energies.}
    \label{parameter-comparison-nspin20-obc}
\end{figure}

A question arises about why to adopt the more computationally expensive QAOA-CD and QAOA-2CD, which require
a minimization of a function depending on $3p$ and $5p$ parameters,
instead of the (at least apparently) more advantageous QAOA-CD-$2p$ and 
QAOA-2CD-$2p$, where the minimization is carried out with respect
to only two parameters per step. To answer this question, we carried out several tests, comparing the two approaches. An example is shown in Fig.~\ref{parameter-comparison-nspin20-obc}, where the residual energy, as a function of the number
of steps, is reported. We take, as a case study, a chain of $N=20$ spins with OBC and uniform couplings ($J_i=1\;\forall i$). 
We see that, at fixed $p$, 
QAOA-CD-$2p$ and QAOA-2CD-$2p$ exhibit the worst performances, thus requiring
a much larger number of iterations to approach the ground-state
energy below a fixed threshold.

\begin{figure}
    (a)\hspace{0.45\columnwidth}(b)\hspace*{0.4\columnwidth}\hfill\\
    \centering
    \includegraphics[width=0.49\columnwidth]{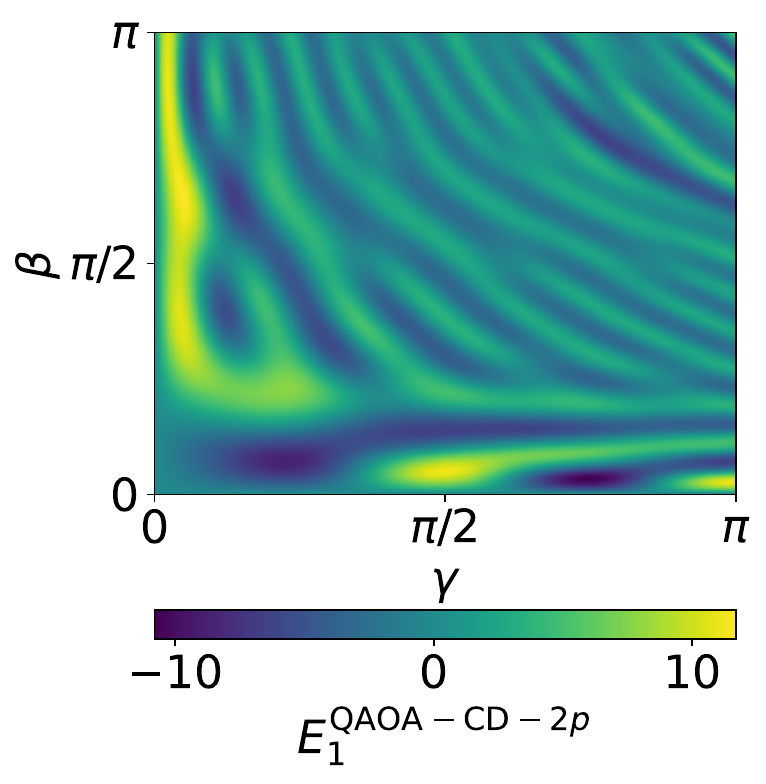}
    \includegraphics[width=0.49\columnwidth]{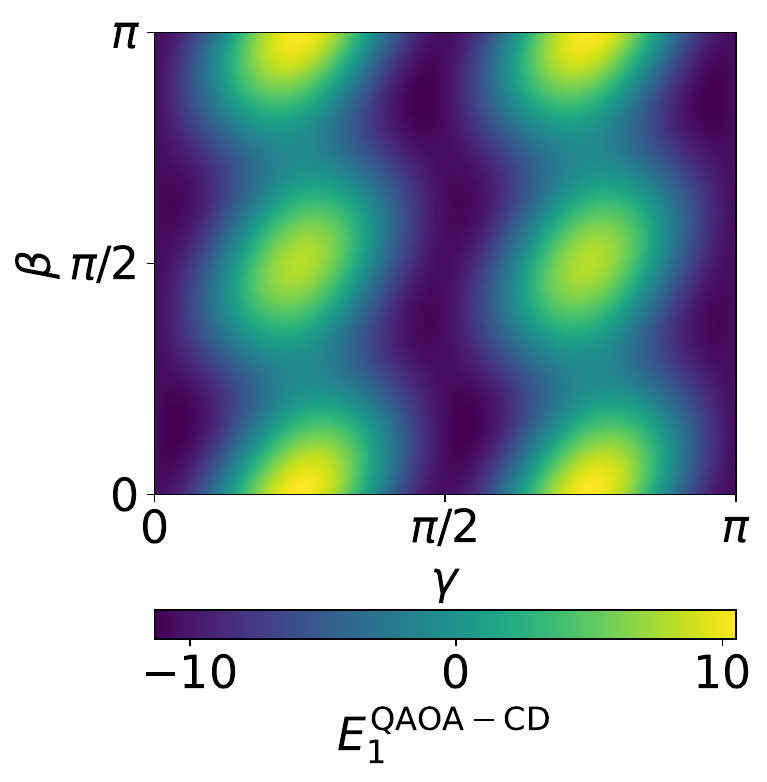}
    \caption{Cost function in the $\beta$-$\gamma$ plane
    at step $p=1$ for a graph of $N=20$ spins with OBC and uniform couplings. Both (a)
    QAOA-CD-2$p$ and (b) QAOA-CD are shown.
    In (b), the cost function depends also on $\alpha$, which has been
    fixed to the value it takes at the minimum. The left panel shows an evidently much more complex landscape. Moreover,
    releasing the constraints of Eqs.~\eqref{eq:no-free-params}
    in the right panel makes the function periodic with respect to
    $\beta$ and $\gamma$, thus restricting the region of the 
    $\beta$-$\gamma$ plane where the minimization has to be carried out.
    }
    \label{cost-QAOA-CD-2p}
\end{figure}

Such an odd behavior is easily explained in terms of cost function
landscapes. It turns out that the cost function of QAOA-CD-2$p$ and  QAOA-2CD-2$p$ have very complicated landscapes, if compared with
QAOA-CD and QAOA-2CD. This is shown in Fig.~\ref{cost-QAOA-CD-2p},
where we compare the cost function for $p=1$ for both the  QAOA-CD-2$p$ and the QAOA-CD for the same systems as that of
Fig.~\ref{parameter-comparison-nspin20-obc}.  For $p=1$ the cost function in the case of the  QAOA-CD-2$p$ depends on the two parameters $(\gamma,\beta)$ while for the QAOA-CD it depends on three independent parameters $(\gamma,\beta,\alpha)$. To show the latter in a
two-dimensional plot we fix $\alpha$ at the value it takes at the minimum of the cost function. Fig.~\ref{cost-QAOA-CD-2p} clearly shows that 
QAOA-CD-2$p$ exhibits a much more complicated energy
landscape if compared with QAOA-CD. The non-periodic sequence of
light and dark regions makes it likely to fall in local minima,
preventing the minimization algorithm to catch or even approach the global minimum.
Moreover, because of the constraints of Eq.~\eqref{eq:no-free-params},
the cost function loses its periodicity with respect to the
$\beta$'s and $\gamma$'s parameters, thus widely enlarging the portion of
the $\beta$-$\gamma$ plane to be explored in the search for the global 
minimum. 

Based on these observations, we will only focus on QAOA-CD and QAOA-2CD in the following.

\subsection{Ring of disagrees}

In this section, we apply QAOA to the ring of disagrees. According to the QAOA, the functional to be minimized is 
obtained from  Eqs.~\eqref{HT-ring} and \eqref{fun-QAOA} as 
\begin{equation}
 \begin{split}
     E_p(\vec \gamma, \vec \beta) =  \left\langle\psi^{(p)}_{\vec \gamma,\vec \beta} \right | \sum_{i=1}^N \sigma_i^Z \sigma_{i+1}^Z\left|\psi^{(p)}_{\vec \gamma,\vec \beta}\right \rangle=\;\\=\sum_{i=1}^N\left\langle\psi^{(p)}_{\vec \gamma,\vec \beta} \right |  \sigma_i^Z \sigma_{i+1}^Z\left|\psi^{(p)}_{\vec \gamma,\vec \beta}\right \rangle\;=\sum_{i=1}^N E_p^{(i)}(\vec \gamma, \vec \beta),
 \end{split}
 \label{fun-QAOA-ring}
\end{equation}
where we have defined 
\begin{equation}
   E_p^{(i)}\equiv \left\langle\psi^{(p)}_{\vec \gamma_p,\vec \beta_p} \right |  \sigma_i^Z \sigma_{i+1}^Z\left|\psi^{(p)}_{\vec \gamma_p,\vec \beta_p}\right \rangle.
\label{fun-QAOA-i}
\end{equation}

\begin{figure}
    (a)\hspace{0.45\columnwidth}(b)\hspace*{0.4\columnwidth}\hfill\\
    \centering
    \includegraphics[width=\columnwidth]{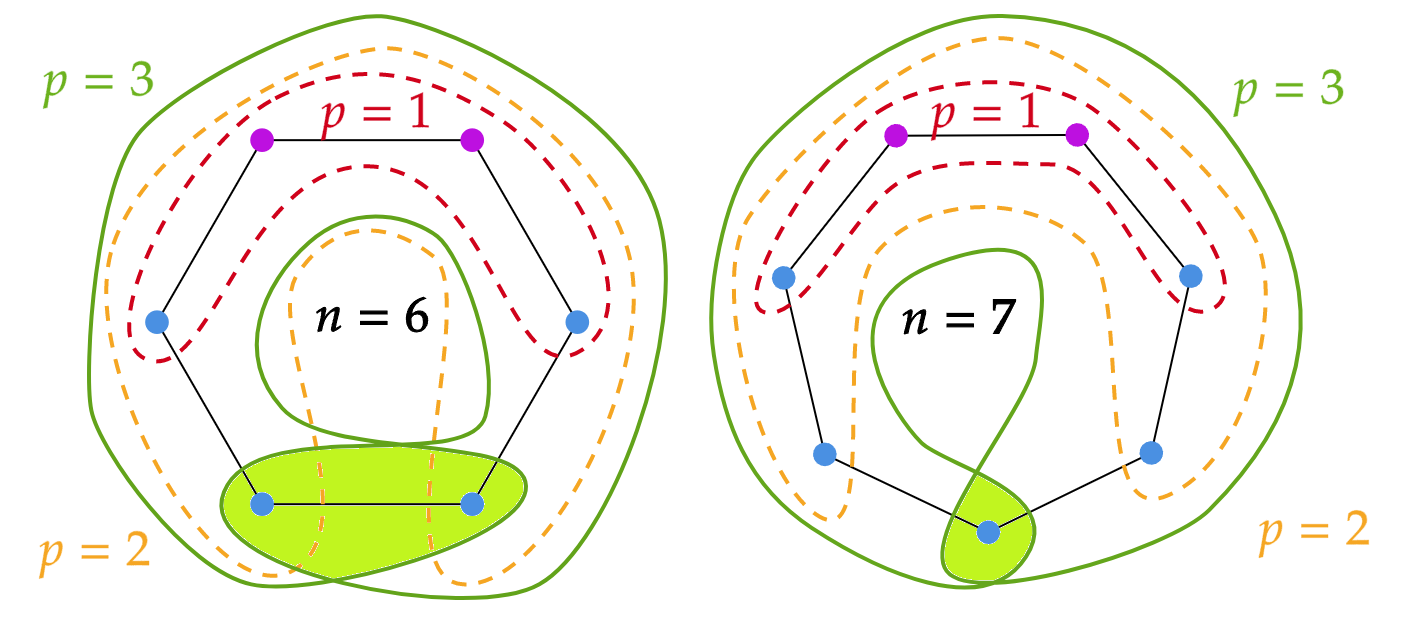}
    \caption{The effective subgraphs at different steps $p$ involved in the calculation of the functional $E_p^{(i)}$
    for the chains composed by $N=6$ (left) and $N=7$ (right) spins, respectively.
    The sites coloured in purple highlight the $i$ and $i+1$ sites. The dashed red, orange and green
    curves enclose the subgraph at steps $p=1,2$ and 3, respectively. For $p=3$ the coloured areas contain the 
    sites that are considered more than once because the subgraph covers the entire original graph.}
    \label{grafo 6-7 spin}
\end{figure}

As detailed in Appendix~\ref{reduced operators}, in Eq.~\eqref{fun-QAOA-i}  only sites within a subgraph, smaller in size 
than $N$ (if $2p+2<N$), are coupled to each other~\cite{farhi2014quantum}. As such, in place of the full Hamiltonian
and trial wave function, we can define the reduced operators and a reduced state
at step $p$ and site $i$ (see Appendix~\ref{reduced operators}) as
\begin{equation}
     \begin{split}
        H_X^{(G'_{p})}=\sum_{j=i-p+1}^{i+p}\sigma_j^X\\H_T^{(G_{p})}=\sum_{j=i-p}^{i+p}\sigma_j^Z\sigma_{j+1}^Z\\\ket{0,G_{p}}=\bigotimes_{j=i-p}^{i+p+1}\ket{+}_j,
    \end{split}
    \label{grafi-ridotti}
\end{equation}
where $G_p$ is the subgraph which the reduced operators act on and $\ket{+}_j$ is an eigenstate of $\sigma_j^X$.
In the previous equations, we omit the dependence on the index $i$ because,
for each site $i$, the translational invariance of the problem ensures that all $G_p$ graphs are equal. 
As $p$ increases, when the subgraph $G_p$ coincides with or contains the graph representing the system, the problem converges to a solution: at this point, the QAOA variational problem contains the number of parameters required to exactly describe the problem and thus to find the exact solution. This is schematically sketched in
Fig.~\ref{grafo 6-7 spin}.

\begin{figure}
    \centering
    \includegraphics[width=0.8\columnwidth]{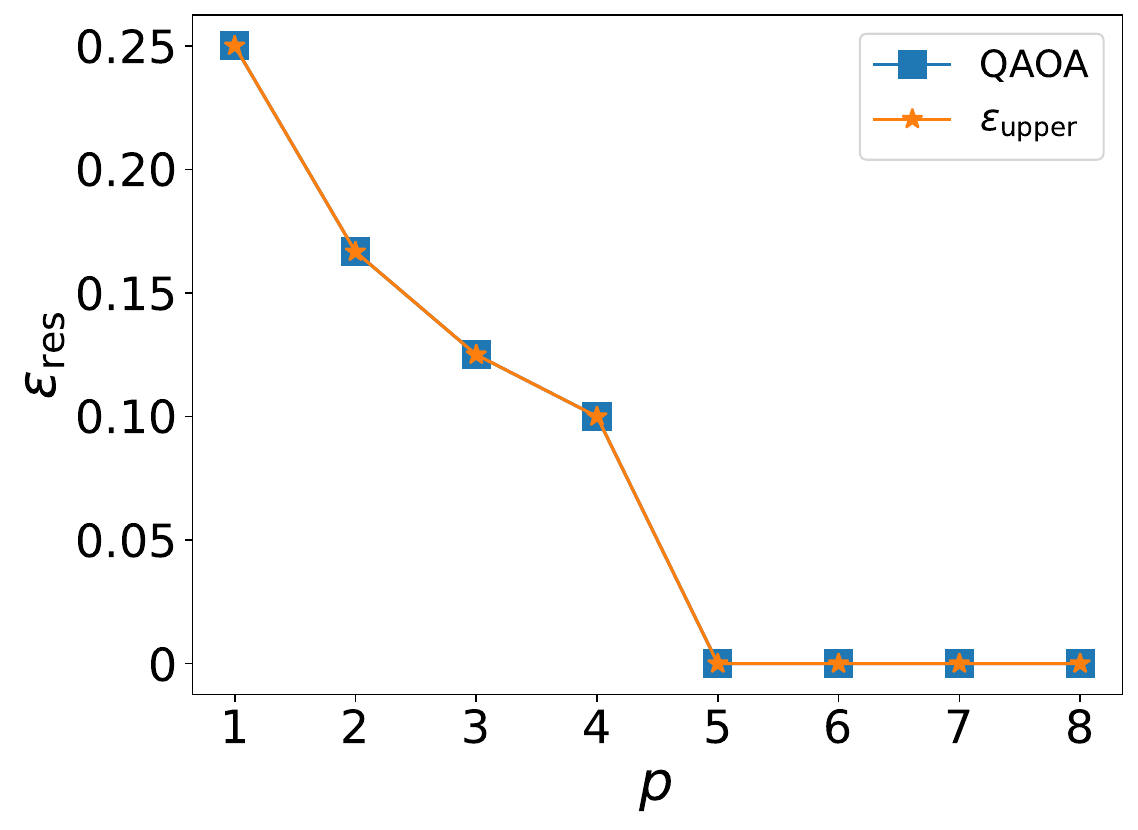}
    \caption{Comparison of the residual energy, $\varepsilon_\text{res}$ (QAOA) (blue curve with squares), and its predicted values according to Eq.~\eqref{eq:eupper}, $\varepsilon_\text{upper}$ (orange curve with stars), for a chain with $N=10$ spins, 
    PBC and uniform couplings $J_i=J=1$ $\forall i$.}
    \label{nspin 6-7}
\end{figure}
Following Ref.~\cite{farhi2014quantum} (see also Appendix~\ref{error}), the partition of the starting graph in subgraphs of dimension $2p+2$ allows  to conjecture
an upper bound  at step $p$ to the residual energy, $\varepsilon^p_\text{upper}$.
It turns out that $\varepsilon^p_\text{upper}=0$ if $2p\ge N$, that is, when the dimension of the subgraph
 exceeds that of the starting graph. On the other hand, if $2p< N$
\begin{equation}
        \varepsilon_\text{upper}^p=
    \begin{cases}
    \frac{1}{2p+2} & \text{ if $N$ is even  } \\
    \frac{N}{N-1}\left(\frac{1}{2p+2}-\frac{1}{N}\right)  & \text{ if $N$ is odd.  } 
    \end{cases}
    \label{eq:eupper}
\end{equation} 
From the previous discussion, we can infer that the number of steps needed to reach full convergence
is $N/2$ ($(N-1)/2$) for even (odd) number of spins. The same conclusion can be drawn also using 
Jordan-Wigner transformations, discussed next.

Following the previous discussion, we can predict the step at which full convergence shows up, i.\,e., 
the first step $p$ that verifies the inequality $2p\ge N$. In Fig.~\ref{nspin 6-7} we compare the residual energy of Eq.~\eqref{eres}, calculated within the QAOA framework, with its conjectured values for a chain of
$N=10$ spins, PBC and uniform couplings. The main result, here,
is that the numerical outcomes fully agree with the theoretical prediction, in particular as far as the step at which full
convergence (zero residual energy) is observed ($p=5$ for the specific investigated system).

\begin{figure}
    (a)\hspace{0.45\columnwidth}(b)\hspace*{0.4\columnwidth}\hfill\\
    \centering   
    \includegraphics[width=0.49\columnwidth]{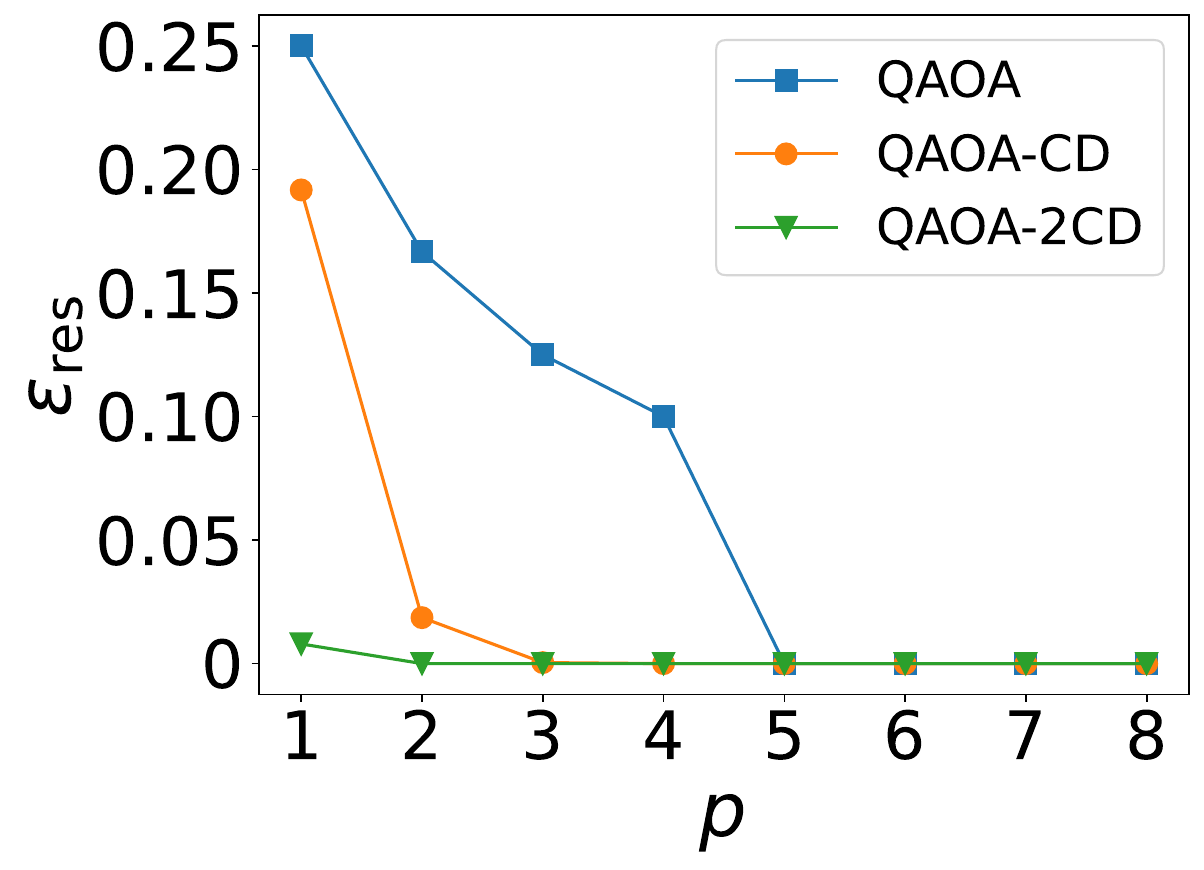}
    \includegraphics[width=0.49\columnwidth]{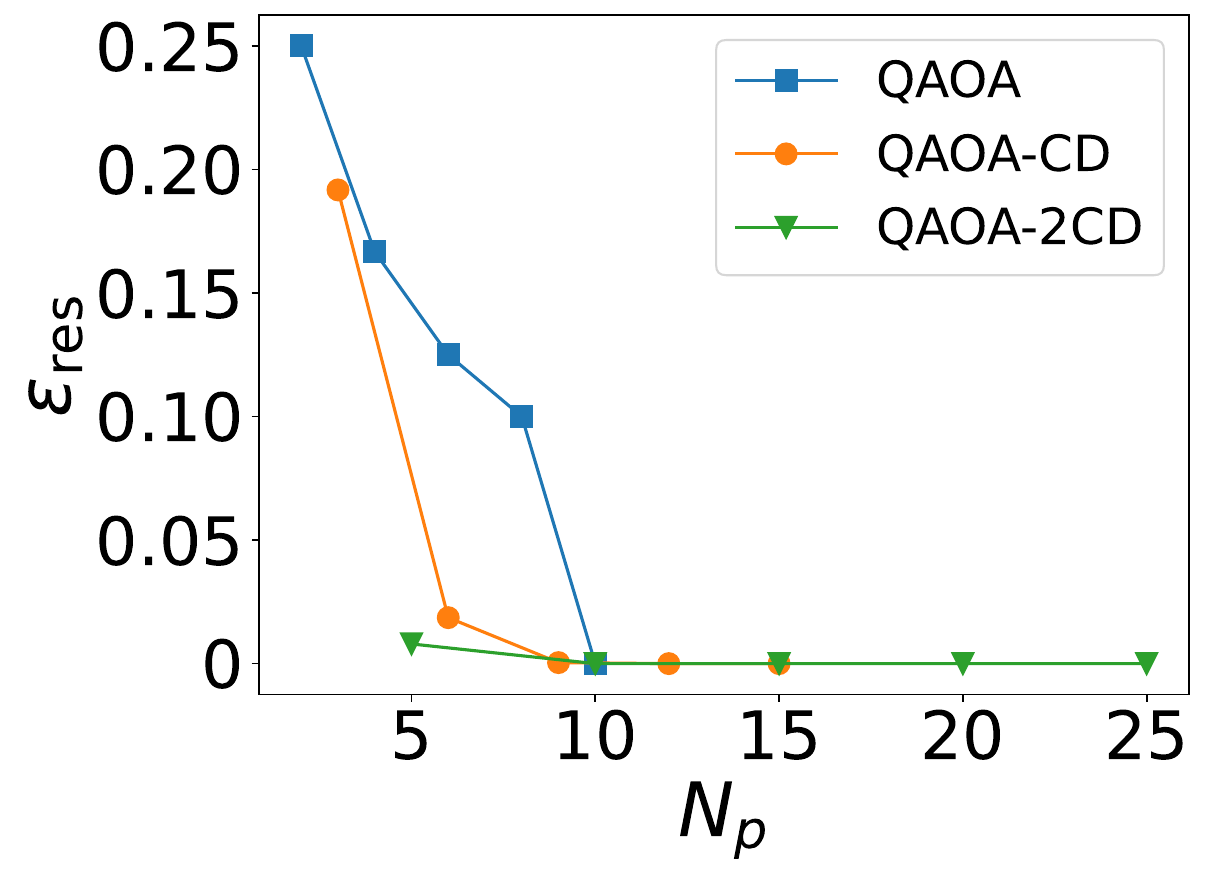}
    \caption{Comparison of the residual energy as obtained within the
    QAOA (blue curve), QAOA+CD (orange curve) and QAOA+2CD (green curve) approaches,
    for a chain of $N=10$ spins and uniform couplings, as a function of the step $p$ (a) and 
    the number of adopted parameters $N_p$ (b).
    We can observe that all algorithms converge with the same number of parameters. }
    \label{step-and-parameter comparison-spin-10-PBC}
\end{figure}

As with the QAOA, in the QAOA-CD and QAOA-2CD approaches, for the ring of disagrees, the involved
operators couple only sites belonging to subgraphs, and thus convergence occurs in a finite number of steps $p$. We can also identify reduced $p$-dependent operators  for these two algorithms. In Fig.~\ref{step-and-parameter comparison-spin-10-PBC} we show the residual energy,
for the same system as that of Fig.~\ref{nspin 6-7}, for all three approaches as a function of both the step $p$
(left panel) and the number of variational parameters involved in the calculation, 
$N_p=p  m$, where $ m=2,3,5$ for QAOA, QAOA-CD and QAOA-2CD respectively.
We observe that the number of steps $p$ required to reach convergence gets lower for the
QAOA-CD and QAOA-2CD algorithms, if compared with conventional QAOA [panel (a)]. Nevertheless, in terms of the number of variational parameters involved in the calculation, it turns out
 that the residual energy goes to zero at the same value of $N_p$ independently of the kind of algorithm used [panel~(b)].

\subsubsection{\label{J-W tranformations}Jordan-Wigner transformations}

A rigorous justification of these observations can be found in terms of Jordan-Wigner (J-W) transformations~\cite{JW}. They allow to express spin operators $\sigma^X$, $\sigma^Y$ and $\sigma^Z$, in terms of spinless fermions such as
\begin{equation}
    \begin{split}
        \sigma_j^X=1-2c^\dagger_jc_j\\\sigma_j^Y=iK_j(c^\dagger_j-c_j)\\\sigma_j^Z=K_j(c_j^\dagger+c_j),
    \end{split}
    \label{operatori JW}
\end{equation}
where
\begin{equation}
    K_j=\prod_{l=1}^{j-1}\left(1-2c_l^\dagger c_l\right).
    \label{Kj}
\end{equation}
$c^\dagger_j$ and $c_j$ are the fermionic creation and annihilation operators. 
These transformations usually map
spin Hamiltonians onto nonlocal fermionic models, due to the presence of the strings $K_j$'s. However,
in the specific case of the Ising model the terms containing the $K_j$'s
cancel out and the Hamiltonian can be expressed in terms of local 
(on-site and nearest neighbors) operators, so we have
\begin{equation}
    \begin{split}
         H_X&=\sum_{j=1}^N\left(1-2c^\dagger_j c_j\right)\\
         H_T&=\sum_{j=1}^N \left(c_j^\dagger c_{j+1} + c_j^\dagger c_{j+1}^\dagger -c_j c^\dagger_{j+1} - c_j c_{j+1}\right).
    \end{split}
    \label{B e C JW}
\end{equation}
Both Hamiltonians do not contain the $K_j$ terms and are bilinear in the fermionic operators. 
The inherent translation invariance allows for a Fourier transform to be carried out on Eq.~\eqref{B e C JW}:
\begin{equation}
    \tilde c_k=\frac{1}{\sqrt{N}}\sum_{j=1}^N e^{ik\omega j}c_j, \;\; \omega=\frac{2\pi}{N}.
\end{equation}
After the transformation, the following basis
\begin{equation}
    \mathbf{\Psi}_k^\dagger=
    \begin{pmatrix}
        \tilde c^\dagger_k & \tilde c_{-k}
    \end{pmatrix}
    \; \;\ \; \;
    \mathbf{\Psi}_k=
    \begin{pmatrix}
        \tilde c_k \\ \tilde c^\dagger_{-k}
    \end{pmatrix}
\end{equation}
 allows rewriting the Hamiltonian, at fixed $k$, in terms of the basis of Pauli operators. It turns out that the resulting Hamiltonian only couples 
 $\tilde c_k$ and $\tilde c^\dagger_{k}$ to $\tilde c_{-k}$ and $\tilde c^\dagger_{-k}$. As a
 consequence, the problem is recast into the solution of a set of
 two-level problems. In particular, for even $N$ we can write
\begin{equation}
\label{eq:single_particle_like}
    \begin{split}
        H_T=\sum_{k=0}^{{\left\lfloor\frac{N-1}{2}\right\rfloor}}\mathbf{\Psi}_k^\dagger H_T^{(k)}\mathbf{\Psi}_k\\H_T^{(k)}=2\hat{\mathbf{k}}\cdot\hat{\mathbf{\sigma}},
    \end{split}
\end{equation}
where $k\in\mathbb{N}$ and
    \begin{align}
        &\hat{\mathbf{k}} = \left( 0, \sin\theta_k, \cos\theta_k \right),\\
        &\theta_k = 2 k \pi / N, \\
        &\hat{\mathbf{\sigma}} = \left( \sigma^X, \sigma^Y, \sigma^Z \right).
    \end{align}
This equation can be easily generalized to odd $N$~\cite{rieffel:2018}. In this way we have transformed the original Hamiltonian $H_T$ describing a ring of $N$ spins into an operator acting on $N/2$ ($(N - 1)/2$ if $N$ is odd~\cite{rieffel:2018}) independent $1/2$-spins, quantized with respect to a new  axis, identified by $\hat{\mathbf{k}}$, within the Bloch sphere. These are referred to as pseudo-spins, to
highlight the difference between the new ``fictitious'' spins and the ``true'' spins of the starting
problem. This transformation is valid for any bilinear and local operator~\cite{rieffel:2018}.

Using this transformation, it is possible to deduce that the number of $N_p$ parameters required for convergence of the QAOA is $N$ for a system with an even number of spins and $N-1$ for a system with an odd number of spins~\cite{rieffel:2018}. The reason is that, for a system made up of a single spin, QAOA converges in a single step to the true ground state. In other words, any point on the surface of the Bloch sphere can be reached with at most two rotations around non collinear axes. Jordan-Wigner transformations map the interacting spin system to a system of noninteracting pseudo-spins in the reciprocal space, and, for each of them, a single QAOA step with just two parameters ensures convergence. Therefore, the number of parameters is $N_p = 2 (N/2)$ ($N_p = 2 [(N-1)/2]$) for even (odd) $N$. This is sketched in Fig.~\ref{grafo 6-7 spin} 
for $6$- and $7$-spin chains, but is valid for all values of $N$.

Higher-order corrections to QAOA are expressed by Eqs.~\eqref{UCD} and~\eqref{eq:u2cd}, which require higher-order commutators $[H_X,H_T]$, $[H_T,[H_X,H_T]]$ and $[H_X,[H_X,H_T]]$. We can also apply  J-W transformation to these terms: 
\begin{equation}
    \left[H_X,H_T\right]=i\sum_{j=1}^N\left(c_j^\dagger c_{j+1}^\dagger +c_j^\dagger c_{j+1} -c_j c_{j+1}^\dagger-c_jc_{j+1} \right),
    \label{commutatore JW}
\end{equation}
and, for QAOA-2CD,
\begin{align}
        &\left[H_X,\left[H_X,H_T\right]\right]\notag\\
        &\quad=8\sum_{j=1}^N \left(c^\dagger_{j-1}c^\dagger_j+c_{j-1}c_j+c_j^\dagger c_{j+1}^\dagger+c_j c_{j+1}\right)
        \label{primo commutatore esplicito JW}
\end{align}
and
\begin{align}
    &\left[H_X,\left[H_T,H_X\right]\right]=-8\sum_{j=1}^N\left(1-2c^\dagger_jc_j+\right.\notag\\
    &\quad\left.+c^\dagger_{j-1}c_{j+1}^\dagger+c_{j-1}^\dagger c_{j+1}+c_{j-1}c^\dagger_{j+1}+c_{j-1}c_{j+1}\right).
    \label{secondo commutatore esplicito JW}
\end{align}
All these terms are bilinear fermionic operators, hence they can be written in the form
\begin{align}\label{eq diag}
    &\left[H_X,H_T\right]\propto \sum_{k'=0}^{\left\lfloor\frac{N-1}{2}\right\rfloor}2\hat{\mathbf{k}}'\cdot\hat{\mathbf{\sigma}},\\
    &s\left[H_X,\left[H_X,H_T\right]\right]-m\left[H_T,\left[H_X,H_T\right]\right]\propto \sum_{k''=0}^{\left\lfloor\frac{N-1}{2}\right\rfloor}2\hat{\mathbf{k}}''\cdot\hat{\mathbf{\sigma}}\notag
\end{align}
for any real numbers $s$ and $m$ and with a suitable choice of the unit vectors $\hat{\mathbf{k}}'$ and
$\hat{\mathbf{k}}''$.

Eq.~\eqref{eq diag} readily generalizes the hypothesis stated for Eq.~\eqref{eq:single_particle_like}, according to which the convergence parameters are $N$ or $N-1$ depending on whether the chain is even or odd. Indeed,
the two Hamiltonians given by Eqs.~\eqref{eq:single_particle_like} and Eq.~\eqref{eq diag} share the same form. 
Hence, even if counterdiabatic terms (first or second order) are added, the same number of parameters $N_p$ of the QAOA is required for convergence in the case of the ring of disagrees. This is consistent with
the results shown in the right panel of Fig.~\ref{step-and-parameter comparison-spin-10-PBC}.

\subsection{Open chain}

Let us now consider a chain with open boundary conditions.
Along the lines of the  PBC problem, it is possible to apply Jordan-Wigner transformations 
also to this system. However, unlike the PBC case, it is not possible to recast the Hamiltonian
in the form of Eq.~\eqref{eq:single_particle_like}, 
since the model of  Eq.~\eqref{h-target-open} is not  translationally invariant. 

\begin{figure}
    (a)\hspace{0.45\columnwidth}(b)\hspace*{0.4\columnwidth}\hfill\\
    \centering   
    \includegraphics[width=0.49\columnwidth]{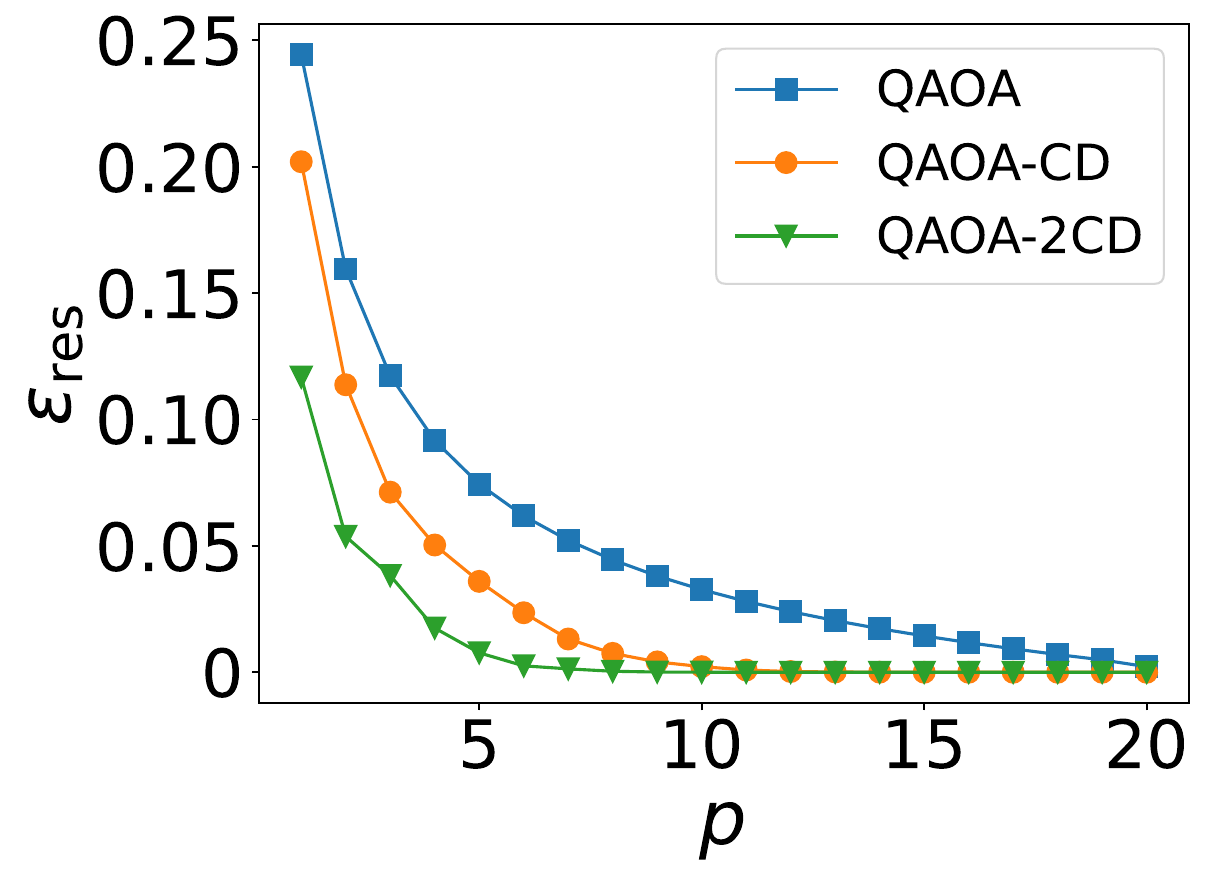}
    \includegraphics[width=0.49\columnwidth]{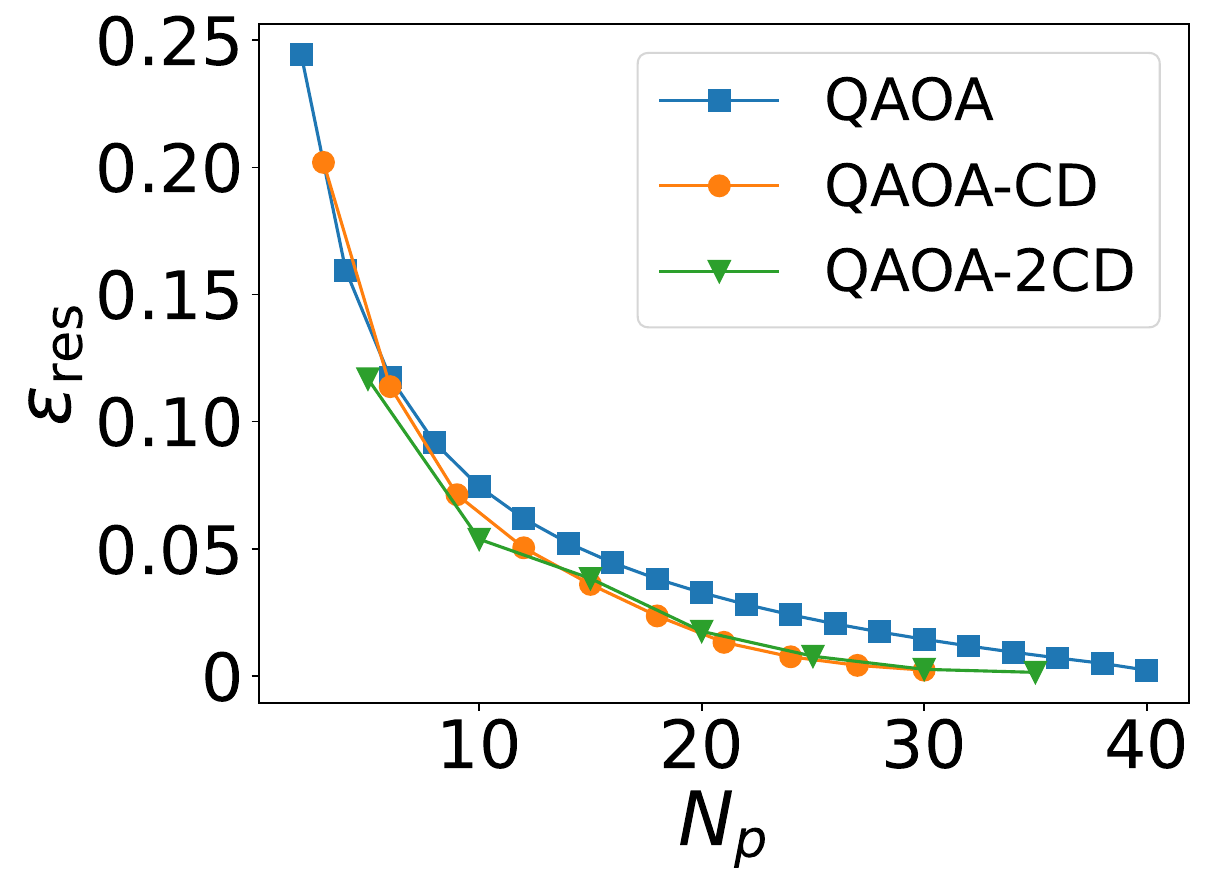}
    \caption{Residual energy for an open chain of $N=20$ spins with uniform couplings as a function of
    (a) the number of steps $p$ and (b) the number of parameters involved in the calculation at the $p$-th
    step, $N_p$ ($N_p=2p,3p,5p$ for QAOA, QAOA-CD, QAOA-2CD, respectively). We see that 
    QAOA-CD and QAOA-2CD converge with fewer steps and fewer parameters than QAOA. }
    \label{step-and-parameter comparison-spin-20-OBC}
\end{figure}

A direct consequence is that it is not possible to infer the number of parameters necessary
for convergence and that there is no analytical approach that can establish \textit{a 
priori} that the number of these parameters must be identical for the three approaches. 
However, the outcomes of our calculations bring out that, as shown in the left panel of
Fig.~\ref{step-and-parameter comparison-spin-20-OBC}, the efficiency in terms of number of steps needed
to reach convergence improves, for the open chain, when moving from QAOA to QAOA-CD to QAOA-2CD
and, at any $p$, we observe $\varepsilon_\text{res}\mbox{(QAOA)} \ge \varepsilon_\text{res}\mbox{(QAOA-CD)} \ge 
\varepsilon_\text{res}\mbox{(QAOA-2CD)}$.
Moreover, despite the lack of analytical evidence, our results indicate that the number of parameters $N_p$ necessary for convergence at a given tolerance remains the same in the three approaches, and also the minimum value of the cost function remains the same even before convergence, as clearly shown in Fig.~\ref{step-and-parameter comparison-spin-20-OBC}(b).

Let us take as an example the QAOA residual energy
at the step $p=20$ ($N_p=40$), $\varepsilon_\text{res}\approx 2.8 \times 10^{-2}$.  
A similar value is reached, within the QAOA-CD approach, at the step $p=10$ ($N_p=30$), whereas
for the QAOA-2CD it occurs at the step $p=6$ ($N_p=30$). Figure~\ref{step-and-parameter comparison-spin-20-OBC}(b) shows that, if we plot the
residual energies as a function of $N_p$, all curves almost collapse onto each other. However,
the QAOA-CD and QAOA-2CD show a slight advantage in the convergence to the solution, with respect to the simple QAOA.
We can conclude that, while QAOA-2CD does not provide any sizable advantage over QAOA-CD in terms of number 
of parameters $N_p$, it converges with fewer steps $p$. For the ring of disagrees it 
has been argued and analytically demonstrated~\cite{rieffel:2018} that the number of the parameters needed to reach convergence is the same for each algorithm that can be expressed in terms of noninteracting pseudo-spins.
On the other hand, in the case of the open chain, this is not an obvious result since the lack of translational invariance does not allow one to write the model in terms of noninteracting pseudo-spins.

\subsection{Random couplings}

In this section we extend our approach to an open chain with random couplings. 
We consider the Hamiltonian of Eq.~\eqref{h-target-random} where $J_i$ is a random variable uniformly distributed in $[-J, J]$ with $J = 1$. We consider spin chains with $N=10,16,20$ spins. For fixed number of spins we analyze $M=20$ instances.  
As expected, the three algorithms behave differently depending on the particular random instance considered, as some instances might be easier to optimize than some others. Therefore, it makes more sense to perform a statistical study to understand the average performances of QAOA and its variants for chains of different  lengths. In particular, we study the residual energy of Eq.~\eqref{eres}. At every step $p$, each instance has different couplings and thus the optimization routines yields different residual energy $\varepsilon_\text{res}$. 

\begin{figure}
    \centering
    \includegraphics[width=0.8\columnwidth]{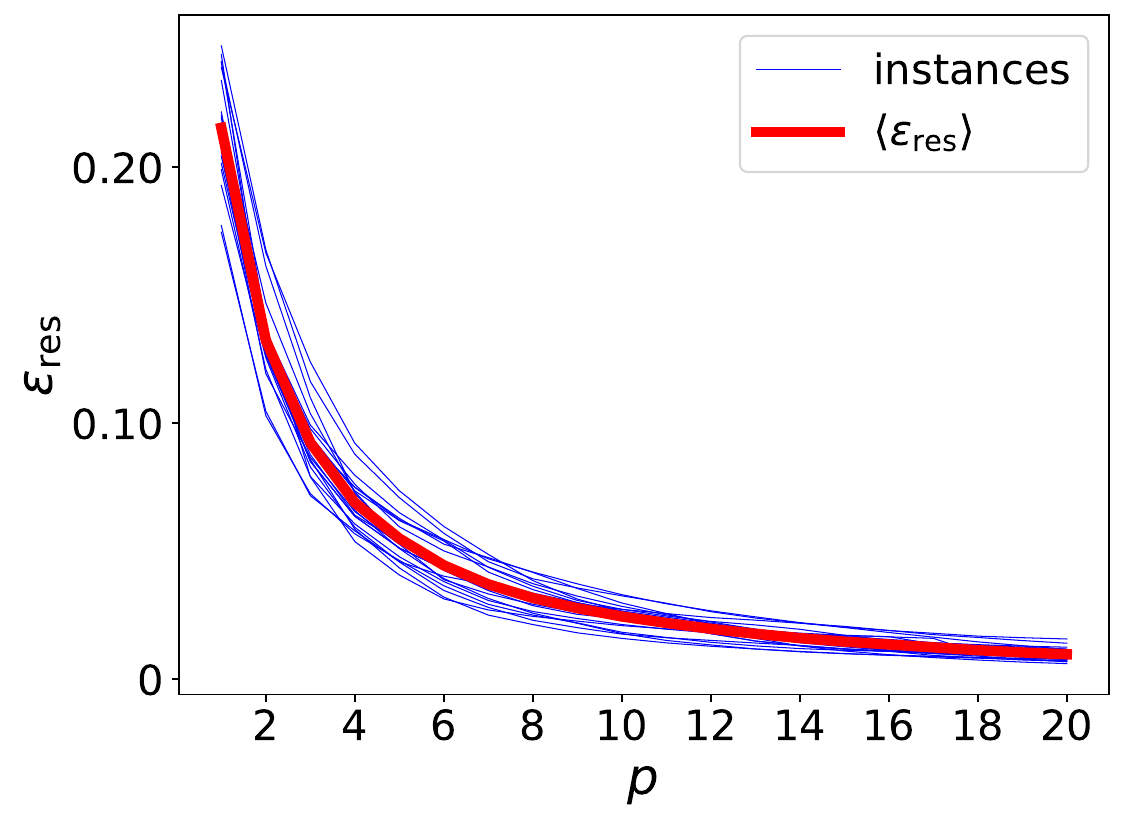}
    \caption{Residual energy of QAOA for $ M = 20$ random instances of a spin chain with $N = 20$ spins and OBC. 
    Here, $\langle\varepsilon_\text{res}\rangle$ is the average of the residual energy over the analyzed random instances. }
    \label{middle-cd0-nspin-20-row1-20}
\end{figure}

First, in Fig.~\ref{middle-cd0-nspin-20-row1-20}, we show our results for standard QAOA. We observe that, due to the simplicity of the cost function, the random instances have low variance with respect to the mean value (shown by the thick red line). When $p$ increases, the residual energy  decreases for all instances, exhibiting a vanishing variance, and the optimization converges to a good solution in all analyzed cases.

\begin{figure}[b]
    (a)\hspace{0.4\columnwidth}(b)\hspace*{0.3\columnwidth}\hfill\\
    \centering
    \includegraphics[width=0.45\columnwidth]{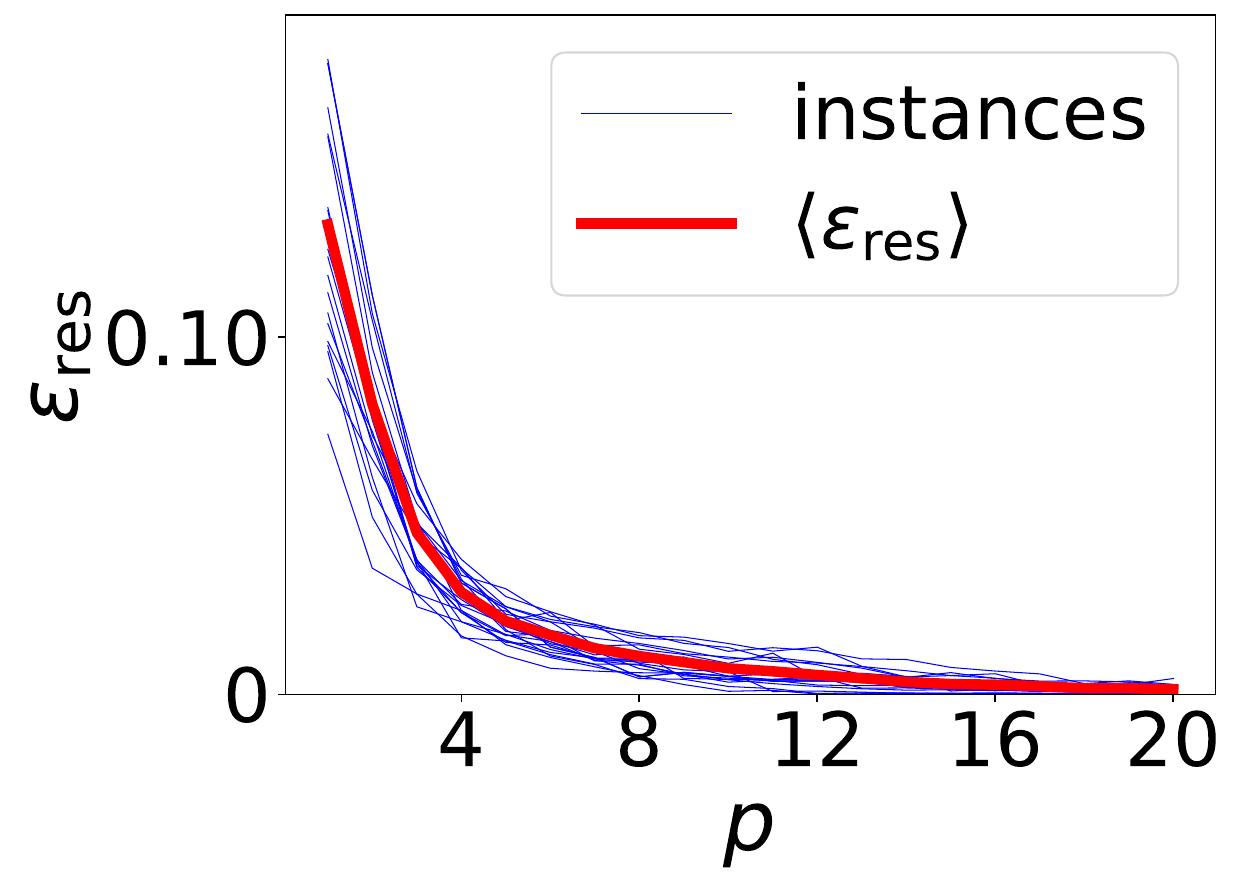}
    \includegraphics[width=0.45\columnwidth]{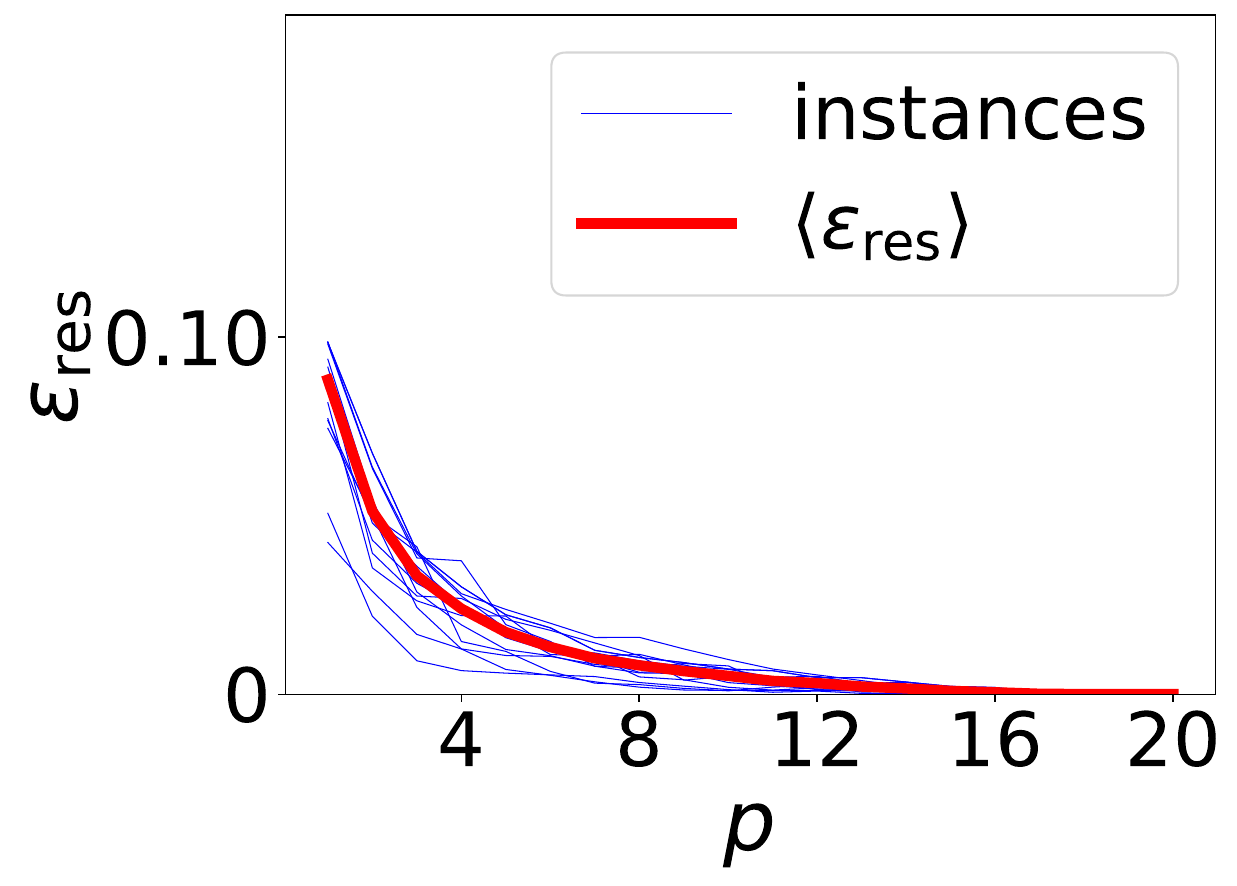}
    \\(c)\hspace{0.4\columnwidth}(d)\hspace*{0.3\columnwidth}\hfill\\
    \centering
    \includegraphics[width=0.45\columnwidth]{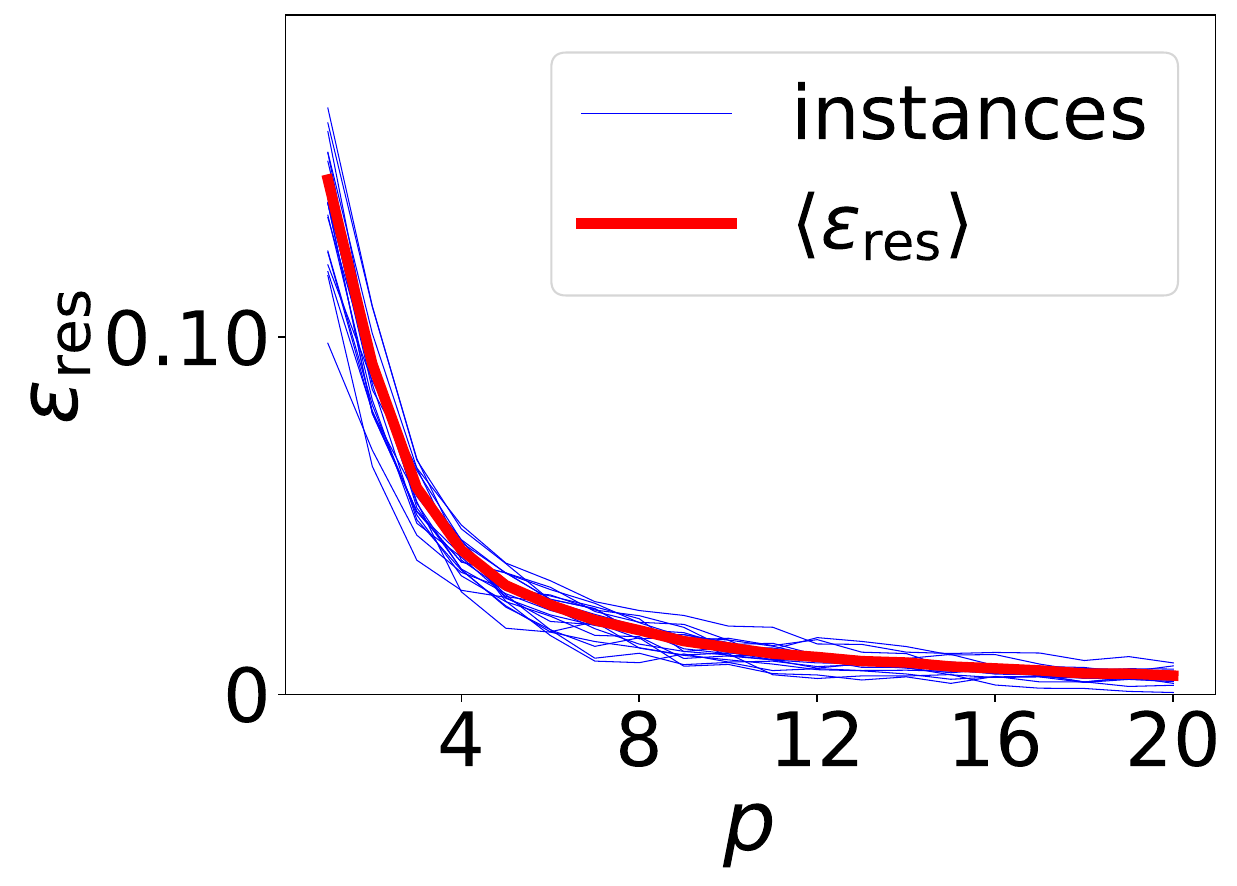}
    \includegraphics[width=0.45\columnwidth]{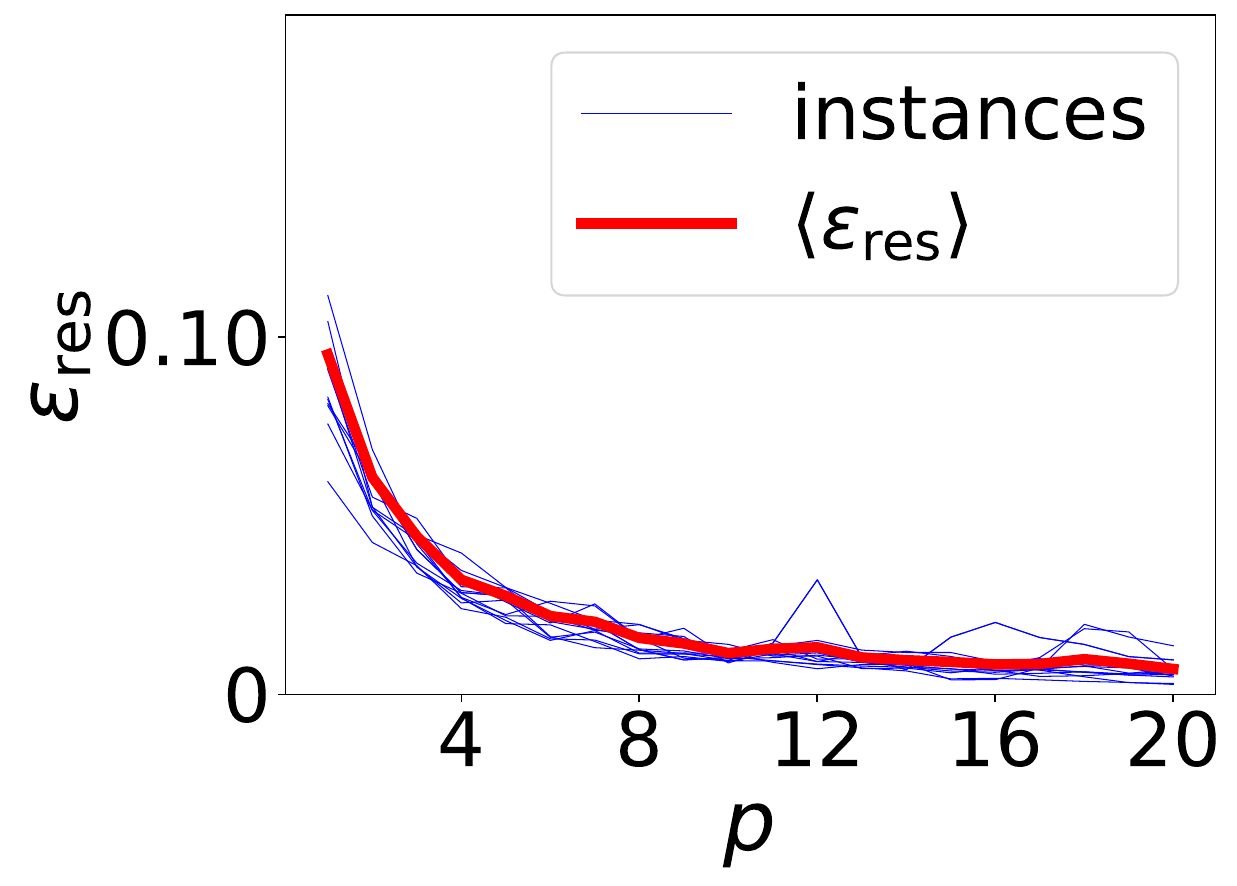}
    \\(e)\hspace{0.4\columnwidth}(f)\hspace*{0.3\columnwidth}\hfill\\
    \centering
    \includegraphics[width=0.45\columnwidth]{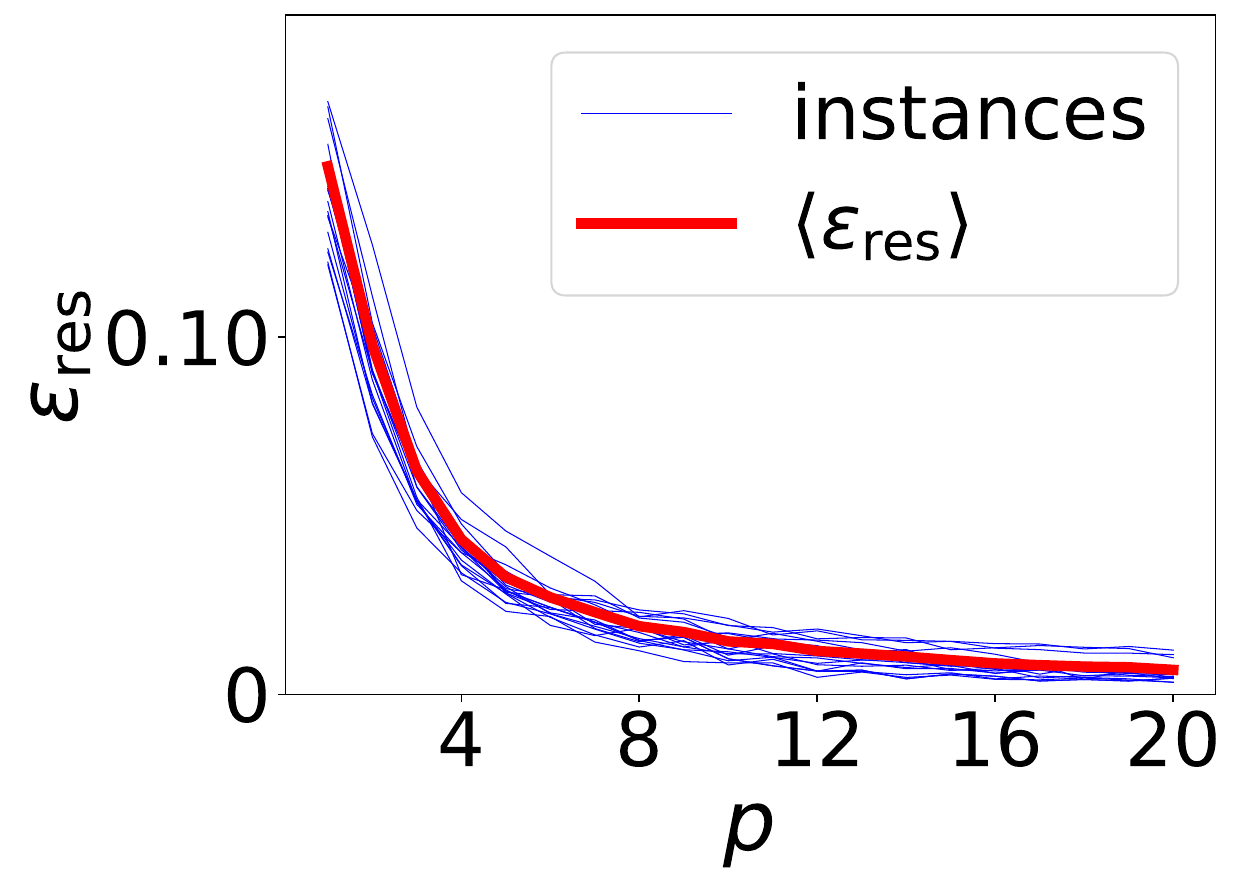}
    \includegraphics[width=0.45\columnwidth]{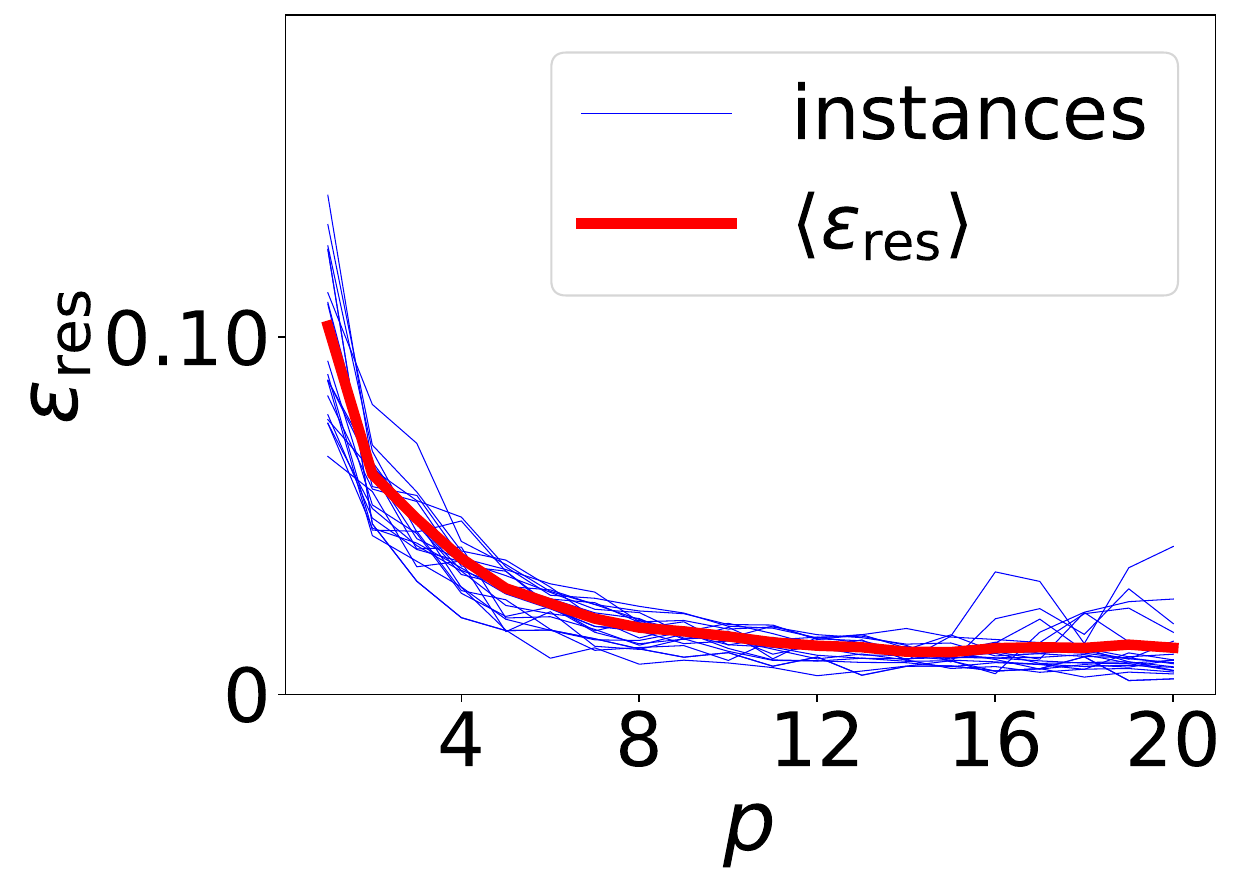}
    \caption{Residual energy $\varepsilon_\text{res}$ of 20 different instances of
    random coupling in an open $N$-spin chain. (a) $N = 10$, QAOA-CD; (b) $N = 10$, QAOA-2CD. (c) $N = 16$, QAOA-CD; (d) $N = 16$, QAOA-2CD. (e)  $N = 20$, QAOA-CD; (f) $N = 20$, QAOA-2CD.}
    \label{CD e CD2 instances}
\end{figure}

\begin{figure*}
    \hspace{-0.2\textwidth}(a)\hspace{0.31\textwidth}(b)\hspace{0.31\textwidth}(c)\hfill\\
    \centering
    \includegraphics[width=0.32\textwidth]{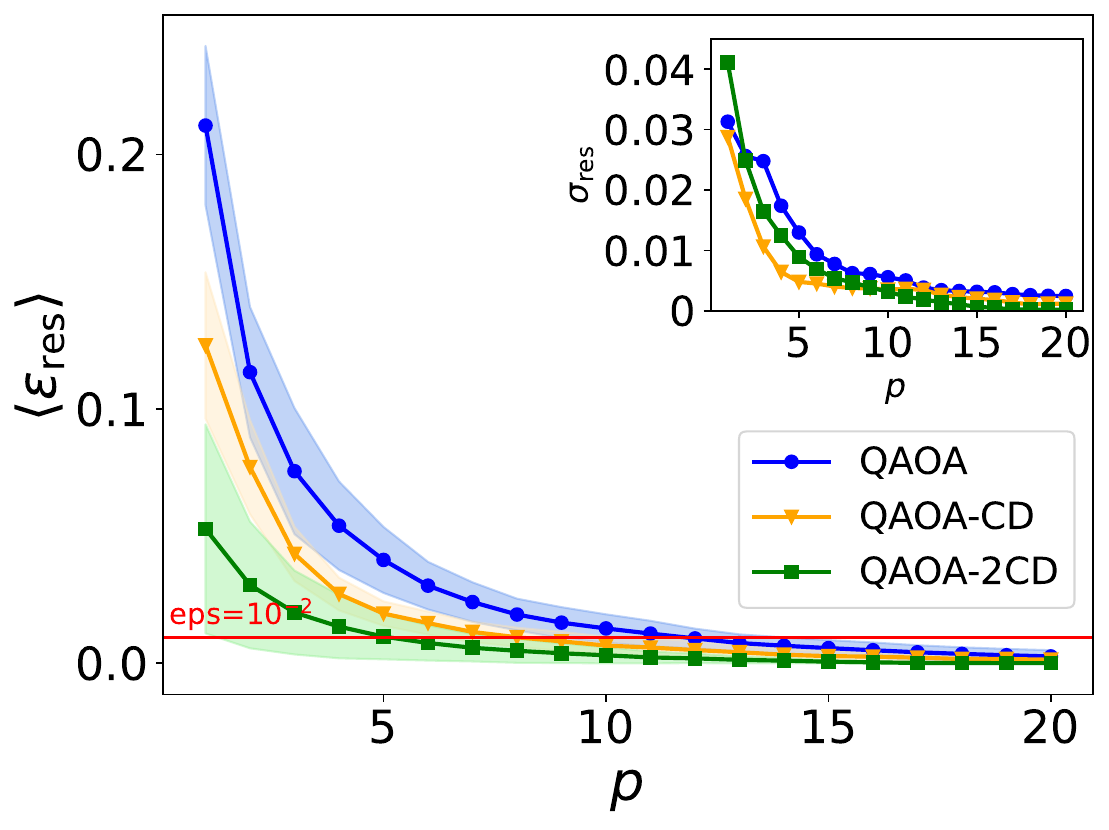}\;
    \includegraphics[width=0.32\textwidth]{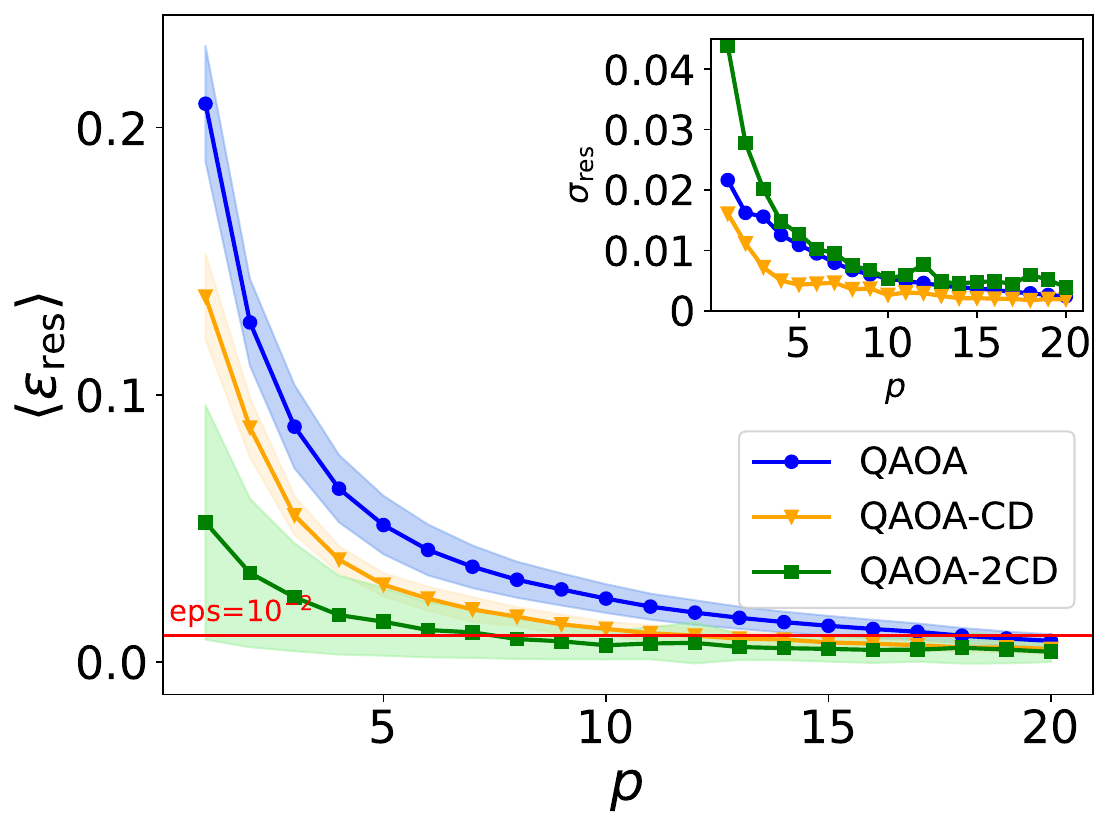}%
    \;
    \includegraphics[width=0.32\textwidth]{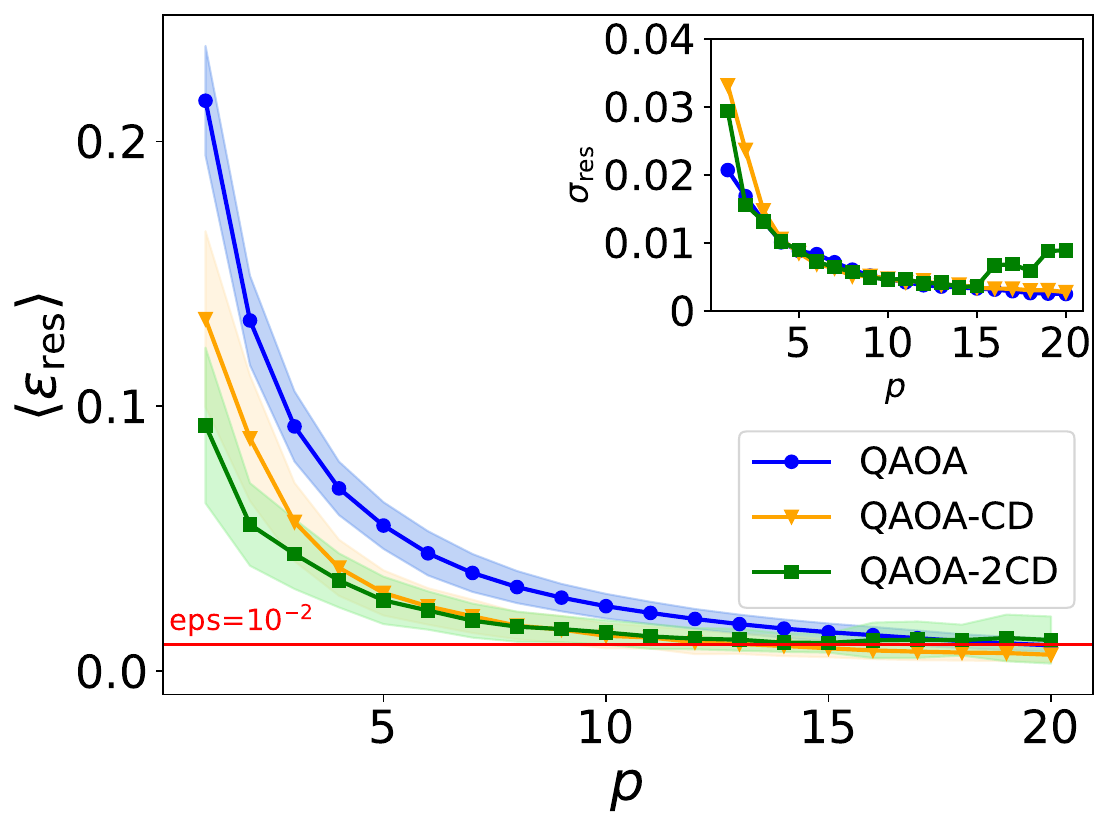}
    \caption{Mean values (solid curves) and standard deviations (shaded regions) of the residual energy over the analyzed random instances of couplings for (a) $N = 10$, (b) $N = 16$, and (c) $N = 20$. In all cases, QAOA-CD and QAOA-2CD converge in fewer steps than QAOA to a value of residual energy below a small, fixed threshold of $10^{-2}$,
    highlighted with the red horizontal line. 
    The insets
    report the standard deviations of the residual energy distribution as a function of the step $p$.}
    \label{midddle-dev-standard}
\end{figure*}

Second, we turn our attention to the counterdiabatically-assisted methods, QAOA-CD and QAOA-2CD. We repeat the optimization for the same $M = 20$ instances for a fair comparison. While running our numerical optimization in this random case, we observed that the INTERP method [see Eq.~\eqref{interpolazione}], which works well for QAOA to select the initial angle for a warm-start of the optimization routine, is not well-suited to this case. The INTERP method allows one to build angles with smooth components and is motivated by the fact that QAOA angles must converge to smooth quantum annealing schedules in the limit $p\to\infty$~\cite{LeoZhou2020}. In order to do so and optimize the angles at step $p$, one has to know the optimal angles at step $(p - 1)$. Ultimately, this means that one needs to know the full optimization history for all the steps $k = 1, \dots, p - 1$, to optimize the angles at step $p$.
However, in the CD case and for the values of $p$ that we have analyzed, we obtained much better results, in terms of residual energy and its variance, by using a brute-force optimization starting from $N_0 = 20$ different initial angles for each value of $p$. On the one hand, in this way we do not enforce any smoothness of the components of the optimized angles. On the other hand, by doing so we are not forced to know the entire optimization history and we can just optimize each QAOA step independently of the previous ones. For this reason, in this case we show our results obtained without INTERP. We summarize our results in Fig.~\ref{CD e CD2 instances}. The different rows correspond to $N = 10$, $16$, $20$, respectively. The first column corresponds to QAOA-CD and the second column to QAOA-2CD. We can see indeed that in all analyzed cases all the residual energies fall close to the mean value and that both methods converge to good approximate solutions within few steps $p$.

\begin{figure}[t]
    \centering
    \includegraphics[width=0.8\columnwidth]{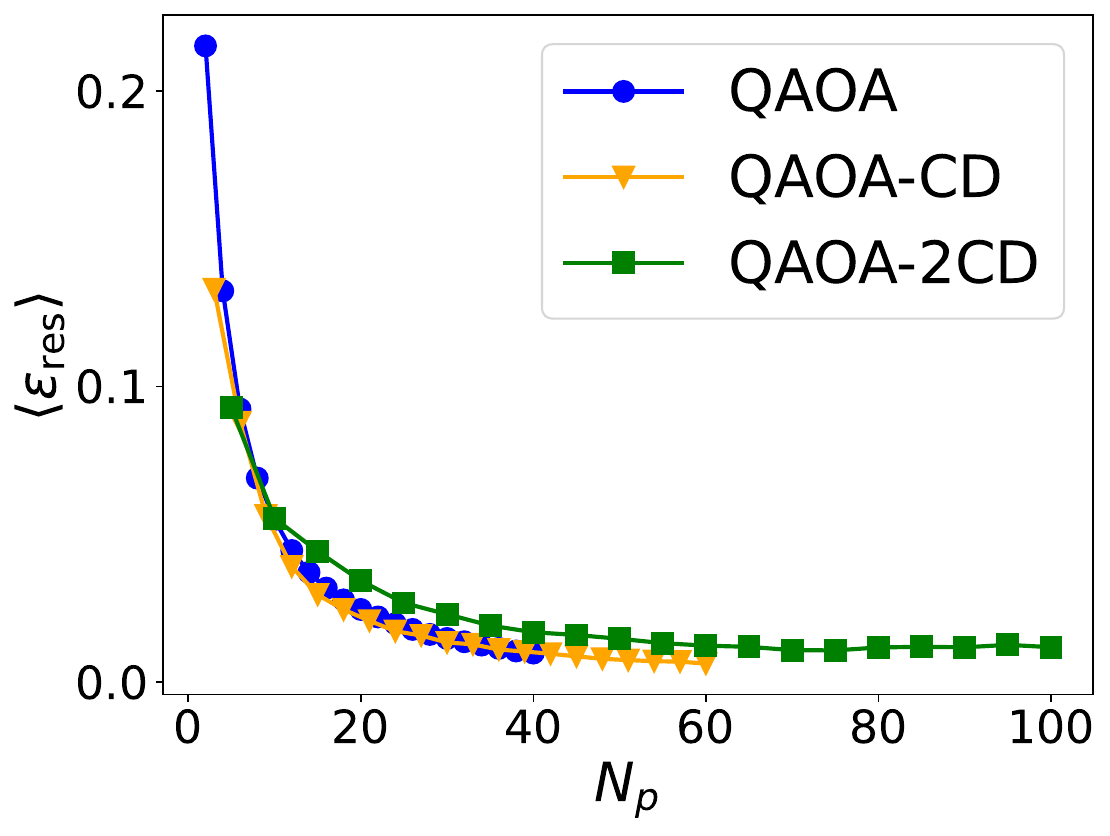}
    \caption{$\varepsilon_\text{res}$ versus  number of parameters $N_p$ for $N = 20$. We can see that the trend of QAOA-CD and QAOA-2CD are similar to that of QAOA.}
    \label{parameters-nspin-20-row1-20-eps-0.001}
\end{figure}

We now compare the average residual energies of all QAOA variants for the analyzed system sizes. We show our results in Fig.~\ref{midddle-dev-standard}, for $N = 10$ [panel~(a)], $N = 16$ [panel~(b)], $N = 20$ [panel~(c)]. In each panel, we show the average residual energy of the three methods using solid lines; the shaded areas around the lines represent values that lie within one standard deviation from the corresponding mean, computed over all the instances at each value of $p$. A plot of the standard deviations as a function of $p$ is also shown in the insets of each panel. At the average level, we observe the same features also found in the uniform case: in particular, we see that QAOA-CD and QAOA-2CD systematically improve QAOA for all values of $p$. Moreover, the standard deviations of the three methods are comparable with each other for all values of $p$ and one order of magnitude smaller than the corresponding mean value.

We can take a look at the average step $p$ required for convergence up to a fixed tolerance in the three methods. For instance, in Fig.~\ref{midddle-dev-standard} we plot the line corresponding to $\varepsilon_\text{res} = 10^{-2}$ and see what the required $p$ for convergence is in the three cases. Here we explicitly comment the case of $N = 10$ but similar results hold for the other sizes as well. We see that, in order to achieve that level of residual energy, QAOA needs $p = 12$ steps, QAOA-CD needs $p = 10$ steps, while QAOA-2CD needs $p = 5$ steps. This fact confirms that adding counterdiabatic corrections to QAOA is able to speed-up convergence, in terms of the number of steps $p$, even in the random case. This is in line with existing results about this subject~\cite{Hegade2022, Chandarana:2022}. {These findings contribute to quantum control research by providing valuable insights and innovative strategies, expanding our knowledge of control techniques in quantum systems~\cite{cepaite2023}.}

Finally, we can study the residual energy as a function of the number of parameters $N_p$, similarly to what we have done in the uniform case. In Fig.~\ref{parameters-nspin-20-row1-20-eps-0.001}, we show our results for $N = 20$. Also in this random case, we see that the average residual energy over all analyzed instances is the same, among the three methods, for any fixed number of parameters $N_p$. {While applicable to the simple case at hand, these results require further investigation to determine their relevance for harder problems~\cite{Zhang2022}.}

\section{Conclusions}\label{sec:conclusions}

In this manuscript we critically assessed the practical advantage of digitized-counterdiabatic QAOA. By investigating two variants of QAOA with first and second-order counterdiabatic corrections, we evaluated their performance and convergence properties in the case of the MaxCut optimization problem.

Our results support the claim that incorporating counterdiabatic unitaries can lead to significant improvements in the efficiency of QAOA in terms of of the step $p$ required for convergence. Indeed, by increasing the complexity of the variational cost function, we observed enhanced convergence towards exact solutions of the optimization problem under consideration. This finding highlights the potential of QAOA-CD and QAOA-2CD to speed-up the optimization process, reducing the required number of steps.

On the other hand, a noteworthy observation from our study is that the total number of variational parameters needed to achieve convergence remains independent of the specific QAOA variant analyzed.  Convergence improvements appear to be significant, however it is important to consider the tradeoff between quantum complexity (i.\,e., the complexity of the quantum circuit that one has to realize) and classical complexity, associated with the more intricate cost function to be minimized. Balancing these complexities becomes crucial when considering the applicability of counterdiabatic techniques in real-world scenarios.

This observation prompts further investigation into the trade-offs and limitations of counterdiabatic approaches in quantum optimization algorithms. Future studies can explore alternative optimization problems domains and evaluate the scalability of counterdiabatic QAOA protocols to larger instances. By understanding the interplay between quantum and classical complexities, we can refine the utilization of counterdiabatic QAOA for a wide range of practical quantum computing applications.

\begin{acknowledgments}
Financial support and computational resources from MUR, PON “Ricerca e Innovazione 2014-2020”, 
under Grant No. PIR01\_00011 - (I.Bi.S.Co.) are acknowledged.
P.L. acknowledges financial support from PNRR
MUR project CN\_00000013-ICSC, PNRR MUR project PE0000023-NQSTI as well as from the project QuantERA II Programme STAQS project that has received funding from the European Union’s Horizon 2020 research and innovation programme under Grant Agreement No 101017733. 

\end{acknowledgments}

\appendix

\section{Open chain with random couplings} \label{Open chain with random couplings}

In this appendix, we analytically compute the cost function of QAOA for the random model with open boundary conditions [Eq.~\eqref{h-target-random}].

We consider Eq.~\eqref{fun-QAOA} for $p=1$ and the initial state given by Eq.~\eqref{start_state}.
For the analytical calculation of the cost function at $p=1$ we start from Eq.~\eqref{fun-QAOA-i}.
Let us focus on the following operator:
\begin{equation}
     e^{i\beta H_X}\sigma_j^Z\sigma_{j+1}^Z e^{-i\beta H_X}.
    \label{op-1}
\end{equation}
We see that all terms without $\sigma^X_j$ or $\sigma^X_{j+1}$ cancel out, so we can re-write 
Eq.~\eqref{op-1} as  
\begin{equation}
     e^{i\beta (\sigma^X_j+\sigma^X_{j+1})}\sigma_j^Z\sigma_{j+1}^Ze^{-i\beta (\sigma^X_j+\sigma^X_{j+1})}.
    \label{op-1-right}
\end{equation}
The same observations apply to the terms in $e^{-i\gamma H_T}$ and its adjoint, so we have (for inner bonds) 
\begin{equation}
    \begin{split}
        e^{i\gamma(J_{j-1}\sigma_{j-1}^Z\sigma_j^Z+J_{j}\sigma_j^Z\sigma_{j+1}^Z+J_{j+1}\sigma_{j+1}^Z\sigma_{j+2}^Z)}\\e^{i\beta (\sigma^X_j+\sigma^X_{j+1})}\sigma_j^Z\sigma_{j+1}^Ze^{-i\beta (\sigma^X_j+\sigma^X_{j+1})}\\e^{-i\gamma(J_{j-1}\sigma_{j-1}^Z\sigma_j^Z+J_{j}\sigma_j^Z\sigma_{j+1}^Z+J_{j+1}\sigma_{j+1}^Z\sigma_{j+2}^Z)}.
    \end{split}
    \label{op-2-right}
\end{equation}
This operator acts nontrivially on the states
\begin{equation}
    \bigotimes_{s=j-1}^{j+2}\ket{+}_s=\ket{+}_{j-1}\otimes\ket{+}_j\otimes\ket{+}_{j+1}\otimes\ket{+}_{j+2}\equiv \ket{++++},
    \label{func-right}
\end{equation}
where $\ket{\pm}_s$ are the eigenstates of $\sigma^x_s$.
To proceed with the calculation, we recall that, if $A$ is a generic operator such that $A^2=\mathbb{1}$,
we can write
\begin{equation}
    e^{i\theta A}= \cos{\theta}+iA\sin{\theta}.
    \label{espansione-exp}
\end{equation}
We can act with every operator on the quantum states of Eq.~\eqref{func-right}:
\begin{equation}
    \begin{split}
        \bigl[\cos(J_{j+1}\gamma)-i\sigma_{j+1}^Z\sigma_{j+2}^Z\sin(J_{j+1}\gamma)\bigr]\ket{++++}=\\=\cos(J_{j+1}\gamma)\ket{++++}-i\sin(J_{j+1}\gamma)\ket{++--},
    \end{split}
    \label{act-1}
\end{equation}
since $\sigma^Z\ket{\pm}=\ket{\mp}$. If we act with $\bigl[\cos(J_{j,j+1}\gamma)-i\sigma_j^Z\sigma_{j+i}^Z\sin(J_{j}\gamma)\bigr]$ on Eq.~\eqref{act-1}, we obtain
\begin{equation}
    \begin{split}
        \cos(\gamma J_{j})\cos(\gamma J_{j+1})\ket{++++}\\-i\cos(\gamma J_{j})\sin(\gamma J_{j+1})\ket{++--}\\-i\sin(\gamma J_{j})\cos(\gamma J_{j+1})\ket{+--+}\\-\sin(\gamma J_{j})\sin(\gamma J_{j+1})\ket{+-+-},
    \end{split}
    \label{act-2}
\end{equation}
and, finally, acting on Eq.~\eqref{act-2} with $\bigl[\cos(J_{j-1}\gamma)-i\sigma_{j-1}^Z\sigma_j^Z\sin(J_{j-1}\gamma)\bigr]$, we have
\begin{equation}
    \begin{split}
        \cos(\gamma J_{j-1})\cos(\gamma J_{j})\cos(\gamma J_{j+1})\ket{++++}\\
    {}-i\cos(\gamma J_{j-1})\cos(\gamma J_{j})\sin(\gamma J_{j+1})\ket{++--}\\
    {}-i\cos(\gamma J_{j-1})\sin(\gamma J_{j})\cos(\gamma J_{j+1})\ket{+--+}\\
    {}-\cos(\gamma J_{j-1})\sin(\gamma J_{j})\sin(\gamma J_{j+1})\ket{+-+-}
    \\{}-i\sin(\gamma J_{j-1})\cos(\gamma J_{j})\cos(\gamma J_{j+1})\ket{--++}
    \\{}-\sin(\gamma J_{j-1})\cos(\gamma J_{j})\sin(\gamma J_{j+1})\ket{----}
    \\{}-\sin(\gamma J_{j-1})\sin(\gamma J_{j})\cos(\gamma J_{j+1})\ket{-+-+}
    \\{}+i\sin(\gamma J_{j-1})\sin(\gamma J_{j})\sin(\beta J_{j+1})\ket{-++-}.
    \end{split}
    \label{act-3}
\end{equation}
At this point, we can act with $e^{-i\beta(\sigma_j^X+\sigma_{j+1}^X)}$.
Remembering that $\ket{+}_s$ is an eigenstate of $\sigma_s^X$, we get
\begin{equation}
    \begin{split}
        e^{-i2\beta}\cos(\gamma J_{j-1})\cos(\gamma J_{j})\cos(\gamma J_{j+1})\ket{++++}\\
    {}-i\cos(\gamma J_{j-1})\cos(\gamma J_{j})\sin(\gamma J_{j+1})\ket{++--}\\
    {}-ie^{i2\beta}\cos(\gamma J_{j-1})\sin(\gamma J_{j})\cos(\gamma J_{j+1})\ket{+--+}\\
    {}-\cos(\gamma J_{j-1})\sin(\gamma J_{j})\sin(\gamma J_{j+1})\ket{+-+-}
    \\{}-i\sin(\gamma J_{j-1})\cos(\gamma J_{j})\cos(\gamma J_{j+1})\ket{--++}
    \\{}-e^{i2\beta}\sin(\gamma J_{j-1})\cos(\gamma J_{j})\sin(\gamma J_{j+1})\ket{----}
    \\{}-\sin(\gamma J_{j-1})\sin(\gamma J_{j})\cos(\gamma J_{j+1})\ket{-+-+}
    \\{}+ie^{-i2\beta}\sin(\gamma J_{j-1})\sin(\gamma J_{j})\sin(\gamma J_{j+1})\ket{-++-}.
    \end{split}
    \label{act-sigmax}
\end{equation}
Now we act with $J_{j}\sigma_j^Z\sigma_{j+1}^Z$ and we obtain
\begin{equation}
    \begin{split}
        e^{-i2\beta}\cos(\gamma J_{j-1})\cos(\gamma J_{j})\cos(\gamma J_{j+1})\ket{+--+}\\
    {}-i\cos(\gamma J_{j-1})\cos(\gamma J_{j})\sin(\gamma J_{j+1})\ket{+-+-}\\
    {}-ie^{i2\beta}\cos(\gamma J_{j-1})\sin(\gamma J_{j})\cos(\gamma J_{j+1})\ket{++++}\\
    {}-\cos(\gamma J_{j-1})\sin(\gamma J_{j})\sin(\gamma J_{j+1})\ket{++--}
    \\{}-i\sin(\gamma J_{j-1})\cos(\gamma J_{j})\cos(\gamma J_{j+1})\ket{-+-+}
    \\{}-e^{i2\beta}\sin(\gamma J_{j-1})\cos(\gamma J_{j})\sin(\gamma J_{j+1})\ket{-++-}
    \\{}-\sin(\gamma J_{j-1})\sin(\gamma J_{j})\cos(\gamma J_{j+1})\ket{--++}
    \\{}+ie^{-i2\beta}\sin(\gamma J_{j-1})\sin(\gamma J_{j})\sin(\gamma J_{j+1})\ket{----}.
    \end{split}
    \label{act-sigma-hamiltonian}
\end{equation}
By combining the previous equations, we are now able to calculate $E_1^{(j)}(\beta,\gamma)$:
\begin{align}\label{fi-calc}
    &E_1^{(j)}(\beta,\gamma) = -\sin(4\beta)\sin(2\gamma J_{j}) \notag\\
    &\quad {} \times\bigl[1-\sin^2(\gamma J_{j-1})-\sin^2(\gamma J_{j+1})\bigr].
\end{align}
Eq.~\eqref{fi-calc} is true for inner sites $j$. For boundary sites, $1$ and $N$, Eq.~\eqref{op-2-right} and Eq.~\eqref{func-right} change: for $j=1$, $\sigma^Z_{j-1}\sigma^Z_j=0$ and, for $j=N$, $\sigma^Z_{j+1}\sigma^Z_{j+2}=0$, so we have 
\begin{align}\label{f1-fN}
    E_1^{(1)}(\beta, \gamma)&=-J_{1}\sin(4\beta)\cos^2(\gamma J_{2})\sin(2\gamma J_{1})\notag\\
    E_1^{(N)}(\beta, \gamma)&=-J_{N-1}\sin(4\beta)\cos^2(\gamma J_{N-2})\sin(2\gamma J_{N-1}).
\end{align}
Finally, we can write
\begin{align}\label{f-anal}
    E_1(\beta, \gamma) &=-\sin(4\beta)\biggl[\sum_{j=2}^{N-2}J_{j}\sin(2\gamma J_{j})\notag\\
    &\quad{}\times\bigl[1-\sin^2(\gamma J_{j-1})-\sin^2(\gamma J_{j+1})\bigr]\notag\\
    &\quad{}+J_{1}\cos^2(\gamma J_{2})\sin(2\gamma J_{1})\notag\\
    &\quad{}+J_{N-1}\cos^2(\gamma J_{N-2})\sin(2\gamma J_{N-1})\biggr].
\end{align}
In particular, if $J_{j}=1\;\forall j$ (uniform couplings), we have
\begin{gather}
    E_1(\beta, \gamma)=\notag \\=-\sin(4\beta)\biggl\{\sum_{i=2}^{N-2}\Big[\sin(2\gamma)\cos(2\gamma)\Big] +2\cos^2{\gamma}\sin{2\gamma}\biggr\}=\notag \\=-\sin(4\beta)\biggl[(N-2)\sin(2\gamma)\cos(2\gamma) +2\cos^2{\gamma}\sin{2\gamma}\biggr].
    \label{f-j=1}
\end{gather}

\section{QAOA for MaxCut and the ring of disagree}
\label{reduced operators}
In this section we will show how QAOA is applied to the ring of disagrees, Eq.~\eqref{HT-ring}. 
Upon analyzing Eq.~\eqref{fun-QAOA-i} for a fixed step $p$ and fixed index $j$, we observe that the calculation of the expected value may not involve all sites and edges. Specifically, there are cases where the sites and edges form a subgraph which is contained within the original graph. This subgraph is denoted as $G_p$.

Let us consider a ring of disagrees with $N$ sites. $\Bar{C}$ as
To derive the general expression of $G_p$ and the expression in Eq.~\eqref{grafi-ridotti}, we start with the case  $p=1$ and follow the discussion at the beginning of Appendix~\ref{Open chain with random couplings}.

In Eq.~\eqref{op-2-right}, valid for $p=1$, we see that the involved spin operators 
are associated with the sites $j-1$, $j$, $j+1$ and $j+2$.
\begin{figure}
    \centering
    \includegraphics[width=\columnwidth]{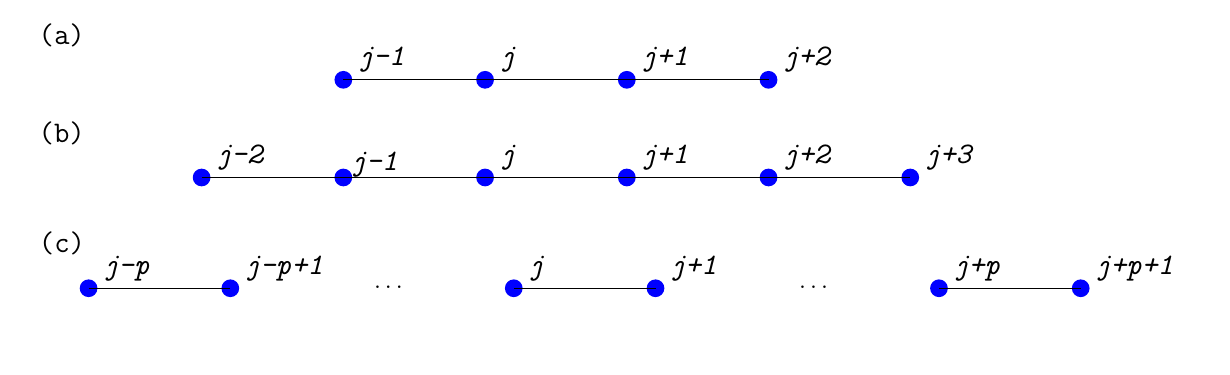}
    \caption{(a) Representation of a subgraph at the step $p=1$. (b) Representation of a subgraph at the step $p=2$. (c) Representation of a subgraph for a generic $p$. }
    \label{figures}
\end{figure}
This allows us to consider subgraphs like the one shown in Fig.~\ref{figures}(a).
The vertices $j$ and $j+1$ are connected with each other and with only one other vertex each: the subgraph considered has $4$ vertices and $3$ edges and has no periodic boundary conditions. We call this subgraph $G_1$.

It is necessary to emphasize that, while the operator $H_T$ of Eq.~\eqref{HT-ring} at step $p=1$ involves
the graph $G_1$, the operator $H_X$ of Eq.~\eqref{hx}, at step $p=1$, involves a graph, which we call $G'_1$, 
composed by only the sites $j$ and $j+1$: this is evident because $H_X$ is a one-body operator. Therefore, the reduced operator of $H_X$ will be
\begin{equation}
    H_X^{(G'_{1})}=\sum_{s=j}^{j+1}\sigma_s^X.
    \label{B ridotto ring}
\end{equation}
In Eq.~\eqref{op-1-right} we see that the only term remaining after the explicit calculation of Eq. \eqref{op-1} are the terms with $\sigma_i^X$ and $\sigma_{i+1}^X$: Eq \eqref{B ridotto ring} in this sense is a generalization at $p>1$ of the term $\exp(-i\beta H_X)$ in Eq. \eqref{op-1-right}. So, in general, we call $G'_{1}$ the subgraph which the operator in Eq. \eqref{B ridotto ring} acts on. 
The QAOA operator $U_{H_X}(\beta)$ thus becomes
\begin{equation}
    U_{H_X^{(G'_{1})}}(\beta)=e^{-i\beta H_X^{(G'_{1})}}.
    \label{UB ridotto ring}
\end{equation}
Considering  Eq.~\eqref{op-2-right}, the reduced operator of $H_T$ instead becomes 
\begin{equation}
    H_T^{(G_1)}=\sum_{s=j-1}^{j+1}\sigma_s^Z\sigma_{s+1}^Z.
    \label{C ridotto ring}
\end{equation}
In this case, the graph to be considered is instead the one depicted above because the terms that do not commute with the reduced operator are those that have $\sigma_j^z$ and $\sigma_{j+1}^z$ and thus
\begin{equation}
    U_{H_T^{(G_1)}}(\gamma)=e^{-i\gamma H_T^{(G_1)}}.
    \label{UC ridotto ring}
\end{equation}
Finally, the initial state becomes
\begin{equation}
    \ket{0,G_{1}}=\bigotimes_{s=j-1}^{j+2}\ket{+}_s,
    \label{stato ridotto ring}
\end{equation}
given by the tensor product of the states involving only the sites associated with $G_1$. 

For $p=2$, instead, we have to consider the subgraph $G_2$ [see Fig.~\ref{figures}(b)],
while the graph $G'_2$ involves the sites $j-1$, $j$, $j+1$, $j+2$, which coincides with the graph $G_1$. It is clear that
the number of involved sites increases with $p$.

For a generic $p$ the subgraph considered becomes like in Fig.~\ref{figures}(c). 
So, for $p>1$, $G'_{p}\equiv G_{p-1}$. The operators are then written as in Eq.~\eqref{grafi-ridotti}, from which the QAOA operators can be calculated.

Thanks to the translational invariance of the model, the expectation value $E_p^{(j)}$ [see Eq.~\eqref{fun-QAOA-i}]
at fixed $p$ and $j$ is the same whatever the pair of neighbouring sites considered. Moreover, for this type of graph, it has been conjectured~\cite{farhi2014quantum,rieffel:2018} that the minimum of the cost function is
\begin{equation}
    M_{p,j}=-\frac{p}{p+1},
    \label{congettura i}
\end{equation}
independently of $j$. Therefore, for the whole ring, the minimum of the cost function reads
\begin{equation}
    M_p=-N\frac{p}{p+1}.
\end{equation}

We now want to verify Eq.~\eqref{congettura i} at least for the case $p=1$. The states and unit operators can be explicitly written as
\begin{equation}
    \begin{split}
        U_{H_X^{(G'_{1})}}(\beta)=e^{-\beta(\sigma_i^X+\sigma_{i+1}^X)}\\U_{H_T^{(G_1)}}(\gamma)=e^{-\frac{\gamma}{2}(3-\sigma_{i-1}^Z\sigma_i^Z-\sigma_i^Z\sigma_{i+1}^Z-\sigma_{i+1}^Z\sigma_{i+2}^Z)}\\\ket{0,G_{1}}=\ket{+}_{i-1}\otimes\ket{+}_i\otimes\ket{+}_{i+1}\otimes\ket{+}_{i+2}.
    \end{split}
    \label{espliciti p1}
\end{equation}
By using Eq.~\eqref{espansione-exp}, we obtain
\begin{equation}
    E_1(\gamma,\beta)=-\frac{1}{2}\sin(4\beta)\sin(4\gamma),
    \label{congettura f p=1}
\end{equation}
which has its minimum at
\begin{equation}
    \begin{split}
        \Bar{\gamma}_1=\frac{\pi}{8}\;\;\text{and}\;\;\ \Bar{\beta}_1=\frac{\pi}{8}\;\;\;\;\text{ or }\;\;\;\;\ \Bar{\gamma}_2=\frac{3}{8}\pi\;\;\text{and}\;\;\Bar{\beta}_2=\frac{3}{8}\pi,
    \end{split}
    \label{soluzioni angoli p=1}
\end{equation}
where the minimum is
\begin{equation}
    M_1=-\frac{1}{2},
    \label{M1}
\end{equation}
in line with Eq.~\eqref{congettura i}.

Based on this argument, it is also possible to predict the step $p^*$ at which the QAOA converges. In fact, convergence is achieved when the subgraph $G_p$ becomes bigger than the entire spin chain~\cite{farhi2014quantum}.
Thus, in general, for an even-spin chain, QAOA converges when $2p^*=N$, so when $p\ge N/2$,
\begin{equation}
    M_{p\geq N/2}=-N.
    \label{maxcut anello pari}
\end{equation}
When $N$ is odd, it is no longer possible to find a value of $p^*$ such that $2p^*=N$. However, the smallest subgraph that covers the entire original spin chain corresponds to $2p^* = (N-1)$, where $M_p=-N+2$. We can conclude that, for a generic ring of $N$ sites  
\begin{equation}
    M_p=\begin{cases}
    -N\frac{p}{p+1} & \text{ for $p<\left\lfloor\frac{N}{2}\right\rfloor$ } \\
    -N & \text{ for even $N$ and $p\geq\left\lfloor\frac{N}{2}\right\rfloor$ } \\
    -N+2 & \text{ for odd $N$ and $p\geq\left\lfloor\frac{N}{2}\right\rfloor$.} 
    \end{cases}
    \label{Mp totale}
\end{equation}

This line of reasoning may be expanded to encompass QAOA-CD and QAOA-2CD, as demonstrated herein.
We start from Eq.~\eqref{fun-QAOA-i}: in standard QAOA, this equation allows us to define the reduced graphs of Eq.~\eqref{grafi-ridotti}. We can extend this study to QAOA-CD and QAOA-2CD. The explicit form of Eq.~\eqref{commutatore JW}, Eq.~\eqref{primo commutatore esplicito JW} and Eq.~\eqref{secondo commutatore esplicito JW} in terms of spin operators is
\begin{align}
    \label{comm 1}&[H_X,H_T]=-2i\sum_{i=1}^N\sigma_i^Y\big(\sigma_{i-1}^Z+\sigma_{i+1}^Z\big)\\
    \label{comm 2} &\Big[H_X,\Big[H_X,H_T\Big]\Big]+\Big[H_T,\Big[H_X,H_T\Big]\Big]\notag \\&= -4\sum_{i=1}^{N}\Big(2\big(\sigma_i^Y\sigma_{i+1}^Y+\sigma_i^Z\sigma_{i+1}^Z\big)-\sigma_i^X\big(1+\sigma_{i-1}^Z\sigma_{i+1}^Z\big)\Big).
\end{align}
To demonstrate that we can restrict the study of Eq.~\eqref{fun-QAOA-i} to subgraphs of the system for QAOA-CD, we observe that 
\begin{gather}
    e^{\alpha[H_X,H_T]}=e^{-2i\alpha\sum_{j=1}^N\sigma_j^Y\big(\sigma_{j-1}^Z+\sigma_{j+1}^Z\big)}=\notag \\=\prod_{j=1}^Ne^{-2i\big(\sigma_{j+1}^Y\sigma_j^Z-\sigma_j^Y\sigma_{j+1}^Z\big)}.
     \label{scomp-QAOA-CD}
\end{gather}
Equation~\eqref{scomp-QAOA-CD} allows us to decompose $U_\text{CD}(\alpha)$ as a product of unitaries acting only on a subgraph and to define a reduced operator also for QAOA-CD:
\begin{equation}
    \Big[H_X,H_T\Big]^{(G''_p)}=-2i\sum_{k=j-2p}^{j+2p+1}\Big(\sigma_{k+1}^Y\sigma_k^Z+\sigma_k^Y\sigma_{k+1}^Z\Big).
    \label{operatore ridotto qaoa-cd}
\end{equation}
We deduce that the subgraph considered in this case has $4p+2$ vertices and $4p+1$ edges.

Similarly, for QAOA-2CD, we can define the reduced operator
\begin{gather}
    \notag\Big(\Big[H_X,\Big[H_X,H_T\Big]\Big]+\Big[H_T,\Big[H_X,H_T\Big]\Big]\Big)^{(G_p^{'''})}=\\ -4\sum_{k=j-3p}^{j+3p+1}\Big(2\big(\sigma_k^Y\sigma_{k+1}^Y+\sigma_k^Z\sigma_{k+1}^Z\big)-\sigma_k^X\big(1+\sigma_{k-1}^Z\sigma_{k+1}^Z\big)\Big).
    \label{operatore ridotto qaoa-2cd}
\end{gather}
Equation~\eqref{operatore ridotto qaoa-2cd} allows us to identify the subgraph consider in QAOA-2CD: a graph with $6p+2$ vertices and $6p+1$ edges.

In Fig.~\ref{step-and-parameter comparison-spin-10-PBC}(a), we see that QAOA converges for a $10$-spin chain at step $p_1=5$, in this case we consider a subgraph of $2p_1+2=12$ vertices that is the first subgraph containing the starting graph. QAOA-CD converges at step $p_2=3$: here, the subgraph has $4p_2+2=14$ vertices. For the same reason, QAOA-2CD converges at step $p_3=2$, its subgraph has $6p_3+2=14$ verices. 
To ensure the validity of our findings, it is important to note that this study holds under the assumption of a periodic chain with boundary conditions. This is due to the translational invariance property that results in all terms in Eq.~\eqref{fun-QAOA-i} being equivalent for every $i$. In the case of an open or disordered chain, this property does not hold, and the results discussed in the main text are nontrivial.

The subgraph study for QAOA-CD and QAOA-2CD is valid also for the ring of disagrees because (only in this case) we can define operators in Eq.~\eqref{operatore ridotto qaoa-cd} and in Eq.~\eqref{operatore ridotto qaoa-2cd}. For the open/disordered chain, we cannot define Eq.~\eqref{operatore ridotto qaoa-cd} and Eq.~\eqref{operatore ridotto qaoa-2cd} because the terms of the sum in Eqs.~\eqref{comm 1} and~\eqref{comm 2} do not commute. Thus, while QAOA is a local algorithm in the sense that it always acts on a part of the system~\cite{Bravyi2020}, the same is not true for QAOA-CD and QAOA-2CD.

\section{Error}
\label{error}

In this section, we discuss the error predicted by the conjecture given by Eq.~\eqref{Mp totale}. Consider $\varepsilon_\text{res}$ for QAOA given by Eq.~\eqref{eres}.
In the case of $H_T$ given by Eq.~\eqref{HT-ring} and even chain, $E_\text{max}=0$, $E_\text{min}=-N$ and $\max_{\vec{\gamma},\vec{\beta}}E_p(\vec{\gamma},\vec{\beta})=M_p$. Then, we have that 
\begin{equation}
    \varepsilon_\text{res}=\varepsilon_\text{res}^p=
    \begin{cases}
    \frac{1}{2p+2} & \text{ if $2p<N$  } \\
    0  & \text{ if $2p\geq N$.  } 
    \end{cases}
    \label{def epsilonmc pari}
\end{equation}

On the other hand, in the case of an odd periodic chain, $E_\text{max}=0$, $E_\text{min}=-N+2$, then
\begin{equation}
    \varepsilon_\text{res}=\varepsilon_\text{res}^p=
    \begin{cases}
    \frac{N}{N-1}\left(\frac{1}{2p+2}-\frac{1}{N}\right) & \text{ if $2p<N$  } \\
    0  & \text{ if $2p> N$.  } 
    \end{cases}
    \label{def epsilonmc dispari}
\end{equation}

%


\end{document}